\documentclass[reprint,aps,prd,superscriptaddress,twocolumn,secnumarabic,nobibnotes,nofootinbib]{revtex4-1}
\usepackage[english]{babel}
\pdfoutput=1
\usepackage[utf8]{inputenc}
\usepackage[normalem]{ulem}
\usepackage{array}
\usepackage[pdftex]{graphicx}
\usepackage{amssymb}
\usepackage{slashed}
\usepackage{braket}
\usepackage{lineno}
\makeatletter
\usepackage[usenames,dvipsnames]{xcolor}

\usepackage{float}

\usepackage{bm}
\usepackage{enumerate}
\usepackage{amsmath}

\usepackage{hyperref}
\hypersetup{colorlinks,citecolor=nicegreen,linkcolor=niceblue}
\hypersetup{colorlinks=true}

\usepackage{pifont}% http://ctan.org/pkg/pifont

\usepackage{multirow}

\allowdisplaybreaks

\newcommand{\nn}{\nonumber}

\newcommand{\e}[1]{\mathrm{\,#1}}       % units
\newcommand{\mc}[1]{\mathcal{#1}}
\newcommand{\mrm}[1]{\mathrm{#1}}

\makeatother

\definecolor{niceblue}{rgb}{0.15,0.15,0.6}
\definecolor{nicegreen}{rgb}{0.1,0.5,0.1}
\definecolor{Red}{rgb}{1.,0.,0.}

\definecolor{Green}{rgb}{0.2,.7,0.2}

\begin{document}
\unitlength = 1mm

\title{Leptoquarks and Real Singlets:\\  A Richer Scalar Sector Behind the Origin of Dark Matter}

\author{Francesco~D'Eramo} \email[Electronic address:]{francesco.deramo@pd.infn.it}
\affiliation{Dipartimento di Fisica e Astronomia, Universit\`a degli Studi di Padova, \\ Via Marzolo 8, 35131 Padova, Italy}
\affiliation{Istituto Nazionale di Fisica Nucleare (INFN), Sezione di Padova, \\ Via Marzolo 8, 35131 Padova, Italy}

\author{Nejc Ko\v snik} \email[Electronic address:]{nejc.kosnik@ijs.si}
\affiliation{Department of Physics, University of Ljubljana, Jadranska 19, 1000 Ljubljana, Slovenia}
\affiliation{Jo\v zef Stefan Institute, Jamova 39, P.\ O.\ Box 3000, 1001
  Ljubljana, Slovenia}

\author{\\Federico Pobbe} \email[Electronic address:]{federico.pobbe@studenti.unipd.it }
\affiliation{Dipartimento di Fisica e Astronomia, Universit\`a degli Studi di Padova, \\ Via Marzolo 8, 35131 Padova, Italy}
\affiliation{Istituto Nazionale di Fisica Nucleare (INFN), Sezione di Padova, \\ Via Marzolo 8, 35131 Padova, Italy}

\author{Aleks Smolkovi\v c} \email[Electronic address:]{aleks.smolkovic@ijs.si}
\affiliation{Jo\v zef Stefan Institute, Jamova 39, P.\ O.\ Box 3000, 1001
  Ljubljana, Slovenia}

\author{Olcyr Sumensari} \email[Electronic address:]{olcyr.sumensari@physik.uzh.ch}
\affiliation{Physik-Institut, Universit\"{a}t Z\"{u}rich, CH-8057, Switzerland}

\preprint{ZU-TH-51/20}  

\begin{abstract}
  We investigate scenarios with $\mathcal{O}(1\e{TeV})$ scalar leptoquarks that act as portals between the Standard Model and Dark Matter. We assume that Dark Matter is a scalar singlet $S$ which couples to a scalar leptoquark $\Delta$ and the Higgs boson via the terms in the scalar potential. In addition, the leptoquark is endowed with Yukawa couplings to quarks and leptons that may address the anomalies in $B$ meson decays. We consider the $SS$ annihilation cross sections to estimate the Dark Matter relic abundance and explore the interplay between astrophysical, collider and flavour physics bounds on such models. In the heavy Dark Matter window, $m_S > m_\Delta$, the leptoquark portal becomes the dominant mechanism to explain the Dark Matter abundance. We find that the leptoquark Yukawa couplings, relevant for quark and lepton flavour physics, are decoupled from the dark matter phenomenology.
By focussing on a scenario with a single leptoquark state, we find that relic density can only be explained when both $\Delta$ and $S$ masses are lighter than $\mathcal{O}(10\e{TeV})$.
\end{abstract}

\maketitle

%\renewcommand{\thefootnote}{\arabic{footnote}}

%\setcounter{footnote}{0}

%\tableofcontents

%\newpage

%%%%%%%%%%%%%%%%%%%%%%
%%%%%%%%%%%%%%%%%%%%%%
%%%%%%%%%%%%%%%%%%%%%%
\section{Introduction}
\label{sec:intro}
%%%%%%%%%%%%%%%%%%%%%%
%%%%%%%%%%%%%%%%%%%%%%
%%%%%%%%%%%%%%%%%%%%%%

Leptoquarks (LQs) are theoretically motivated hypothetical bosonic degrees of freedom that couple at tree-level to quark-lepton pairs~\cite{Buchmuller:1986zs,Dorsner:2016wpm}. They naturally appear in theories unifying quarks and leptons~\cite{Pati:1974yy,Georgi:1974sy} and composite Higgs models~\cite{Schrempp:1984nj,Wudka:1985ef}, and they provide a viable mechanism to explain neutrino masses~\cite{Mahanta:1999xd,Chua:1999si}. Their interactions with fermions also make them good candidates to address phenomenological problems in quark-lepton transitions, such as the discrepancies in $B$-meson decays observed at LHCb and the $B$-factories~\cite{deSimone:2020kwi}, or to address discrepancies in chirality-suppressed observables, such as the anomalous magnetic moment of leptons~\cite{Bennett:2006fi,Abi:2021gix,Aoyama:2020ynm}. Whether these particles exist near the TeV scale remains a question to be answered by current and future experiments. In the meantime, it is natural to ask if LQs could also be related to other open questions in particle physics.

One of the most striking motivations for physics beyond the Standard Model (SM) is the evidence for Dark Matter (DM). Astrophysical and cosmological observations have accumulated indisputable evidence at vastly different length scales~\cite{Bertone:2004pz}. These observations infer the presence of DM through its gravitational effects, but they tell us nothing about its microscopic nature. There are constraints from observations such as precise determination of its relic density~\cite{Aghanim:2018eyx}, strong bounds on its electromagnetic interactions~\cite{McDermott:2010pa,DEramo:2016gqz}, and the fact that it must be stable on cosmological timescales~\cite{Aoyama:2014tga,Slatyer:2016qyl}. Nevertheless, these constraints still leave room for a plethora of particle candidates~\cite{Feng:2010gw}.

An appealing scenario is the one where DM particles interact with the visible world enough to achieve thermal equilibrium in the early universe. As the universe expands and cools down, interactions between DM particles become less frequent. Thermal equilibrium is lost eventually, and such a departure from equilibrium is the physical process that sets the relic density. Remarkably, the resulting value depends only on dark sector masses and couplings that we can test in our experiments~\cite{Lee:1977ua}. 

In this paper, we explore scalar LQs as mediators to the dark sector and whether this type of portal could explain the observed DM properties. We focus on scenarios where the DM particle is a scalar singlet that interacts with one of the scalar LQs via quartic interactions in the scalar potential. This scenario was the focus of Ref.~\cite{Choi:2018stw}. We perform a comprehensive analysis that extends previous studies in several ways. We will consider the impact of all scalar LQ representations and provide the most general expressions for DM phenomenology. Moreover, we will study the interplay between the DM relic abundance with theoretical considerations of stability and perturbativity of the scalar potential, with direct and indirect DM detection constraints, as well as with flavour physics bounds. In particular, we will show that the LQ couplings needed to explain the observed DM relic abundance have little impact in flavour physics phenomenology. Furthermore, we will argue that the LQ and DM masses in the viable models must satisfy $m_\Delta < m_S < \mc{O}(10\,\mrm{TeV})$ in order to comply with the constraints from the stability and perturbativity of the scalar potential.

Several works have recently explored the possible connection between LQs and the observed DM abundance. Besides the possibility of scalar singlet DM considered in this work, the DM particle can be a fermionic singlet with gauge-invariant interactions with specific scalar and vector LQ representations~\cite{Azatov:2018kzb,Mandal:2018czf}. DM can also belong to higher dimensional electroweak multiplets which can couple to the Pati-Salam vector LQ $U_1=(\mathbf{\bar{3},1},2/3)$ \cite{Guadagnoli:2020tlx}, which was proposed as a viable candidate to explain the so-called $B$-physics anomalies~\cite{DiLuzio:2017vat,Bordone:2017bld,Blanke:2018sro}. The connection between DM and the $B$-anomalies has also been explored in the context of composite vector LQ scenarios in Ref.~\cite{Cline:2017aed}.

Our setup and conventions are introduced in Sec.~\ref{sec:lq-classification}. We list the phenomenological constraints on our framework in Sec.~\ref{sec:phenomenology}. In particular, we compute the DM relic density, we impose experimental bounds from DM direct and indirect searches, and we consider constraints from collider physics. We investigate a concrete scenario with DM mass above the weak scale in Sec.~\ref{sec:results}. We discuss the possible connection with flavour anomalies in Sec.~\ref{sec:FA}, and we conclude in Sec.~\ref{sec:conclusion}. Technical details are deferred to appendices.

%%%%%%%%%%%%%%%%%%%%%%
%%%%%%%%%%%%%%%%%%%%%%
%%%%%%%%%%%%%%%%%%%%%%
\section{A Richer Scalar Sector: \\ LQs and DM}
\label{sec:lq-classification}
%%%%%%%%%%%%%%%%%%%%%%
%%%%%%%%%%%%%%%%%%%%%%
%%%%%%%%%%%%%%%%%%%%%%

In this Section we present our framework. We write down Yukawa interactions allowed by gauge invariance for different LQ representations, and scalar potential interactions for the DM candidate. We adopt the notation of Refs.~\cite{Buchmuller:1986zs,Dorsner:2016wpm} and specify LQ states by their SM quantum numbers $(SU(3)_c,SU(2)_L,Y)$. 

We define the covariant derivative
\begin{equation}
  \label{eq:Dmu}
D_\mu = \partial_\mu + i g_1\, Y B_\mu+i g_2\, T^k W_\mu^k+i g_3\, T^A G_\mu^A\,,
\end{equation}
where $Y$ is the hypercharge, and $T^A$ and $T^k$ are the relevant $SU(3)_c$ and $SU(2)_L$ generators, respectively. After the electroweak symmetry breaking, we can write
\begin{align}
  \label{eq:cov-dev}
    D_\mu &= \partial_\mu + i \dfrac{g_2}{\sqrt{2}}\, (T^+ W_\mu^+ +T^- W_\mu^-)\\
    &+i \dfrac{g_2}{c_W}\,  (T_3-Q s_W^2)Z_\mu+i e\, Q A^\mu+i g_3\, T^A G_\mu^A\,,\nonumber
	\end{align}
where we idenfity the electric charge generator $Q=Y+T_3$, the weak isospin raising/lowering matrices $T^\pm=T_1\pm i\, T_2$, and the Weinberg angle $\theta_W$ (we write for shortness $s_W\equiv \sin \theta_W$ and $c_W\equiv \cos \theta_W$). 

\subsection{Yukawa interactions between LQs and SM fermions}

Scalar LQs have Yukawa couplings with the SM quarks and leptons. In order to describe the general features of LQ portals, we provide the general and convenient parameterization 
\begin{align}
\label{eq:yukF0}
\mathcal{L}_{\mathrm{Yuk}} &= \sum_{a,i,j} \overline{q_i} \big{[} y^R_{ij} P_R+y^L_{ij} P_L \big{]} l_j \, \Delta^{(a)}+\mathrm{h.c.}
\end{align}
These interactions are understood to be written for fermion mass eigenstates, and the Yukawa couplings $y^L$ and $y^R$  are matrices in flavour space. The chiral projectors $P_{L,R}$ isolate the corresponding chiralities of the fermion fields they act on. In our notation, $l_i$ stands either for a charged lepton or a neutrino. On the contrary, $q_i$ can be either a quark or its charge-conjugate. Finally, the scalar field $\Delta^{(a)}$ stands for a specific component $(a)$ of the LQ multiplet $\Delta$. 

We introduce concrete realizations for the interactions in Eq.~\eqref{eq:yukF0}. Following Ref.~\cite{Buchmuller:1986zs}, we introduce the fermion number $F$ defined as $F = 3B + L$, with $B$ and $L$ the baryon and lepton numbers, respectively. We assume there are no fermionic SM singlets.

Two scalar LQs can couple to fermion currents not carrying a net fermion number
\begin{widetext}
\begin{align}
\label{eq:yuk-R2}
R_2= \left( \mathbf{3}, \mathbf{2},7/6\right):\qquad \mathcal{L}_{R_2}=& -\left( y_{R_2}^L\right)_{ij}\, \bar{u}_{R_i} R_{2}^T i\tau_2 L_{j} + \left( y_{R_2}^R\right)_{ij} \bar{Q}_i R_{2} \ell_{R_j} +\textrm{h.c.}\,, \\[0.5em]
\label{eq:yuk-R2tilde}
\widetilde R_2= \left( \mathbf{3}, \mathbf{2},1/6\right):\qquad \mathcal{L}_{\widetilde{R}_2}=&- \left(y_{\widetilde{R}_2}^L\right)_{ij}\, \bar{d}_{R_i} \widetilde{R}_2^T  i\tau_2 L_j +\textrm{h.c.}\,.
\end{align}
\end{widetext}
Left-handed fermion doublets in the above expressions are defined as $Q_i=\left[(V^\dagger u_L)_i\; d_{Li}\right]^T$ and $L_i=\left[(U \nu_L)_i\; \ell_{Li}\right]^T$, where $V$ and $U$ denote respectively the Cabibbo–Kobayashi-Maskawa (CKM) and Pontecorvo–Maki–Nakagawa–Sakata (PMNS) matrices. Since neutrino masses are irrelevant for the phenomenology we will discuss, we are free to set $U= \mathbf{1}$. The Yukawa interactions in Eqs.~\eqref{eq:yuk-R2} and \eqref{eq:yuk-R2tilde} conserve the fermion number $F$ as LQs have $F = 0$ in these cases. Baryon and lepton numbers are conserved upon assigning them to the LQ field as well. 

%%%%%%%%%%%%%%%%%%%%%
%\begin{widetext}
\begin{table*}
\renewcommand{\arraystretch}{1.8}
\begin{tabular}{|c|cc|cc|ccc|}
\hline 
\multicolumn{8}{|c|}{ Charged leptons} \\ \hline\hline
Coupling & $q$ & $l$ &   $R_2$  & $\widetilde{R}_2$ & $S_1$ &  $S_3$  & $\widetilde{S_1}$ \\ \hline\hline
 \multirow{2}{*}{$y_{ij}^L$} & $u_i$ & $\ell_j$ &   $-\Big{(}y_{R_2}^L\Big{)}_{ij}$     &  $0$ & $\Big{(}V^\ast y_{S_1}^L\Big{)}_{ij}$     	&  $-\Big{(}V^\ast y_{S_3}^L\Big{)}_{ij}$  & $0$ \\%[0.3em] 
         & $d_i$ & $\ell_j$  &  $0$     & $-\Big{(}y_{{\widetilde{R}_2}}^L\Big{)}_{ij}$     & $0$	& $-\sqrt{2}\Big{(}y_{S_3}^L\Big{)}_{ij}$  & $0$\\%[0.3em]
  \hline
%  \cdashline{1-8}
\multirow{2}{*}{$y_{ij}^R$} & $u_i$ & $\ell_j$  &  $\Big{(}V y_{R_2}^R\Big{)}_{ij}$      &  $0$ & $\Big{(}y_{S_1}^R\Big{)}_{ij}$ 	& $0$ & $0$ \\%[0.3em]  
&$d_i$& $\ell_j$  &    $\Big{(}y_{R_2}^R\Big{)}_{ij}$    &  $0$ & 	$0$ & $0$ & $\Big{(}y_{\widetilde{S}_1}^R\Big{)}_{ij}$ \\
\hline
               \end{tabular}
               \caption{ %\sl \small
                 \label{tab:lq-matching-l} 
                 Expressions for the coefficients $y^L_{ij}$ and $y^R_{ij}$ in Eq.~\eqref{eq:yukF0} for each LQ state listed in Sec.~\ref{sec:lq-classification} interacting with charged leptons. $V$ denotes the CKM matrix and the scalar LQ states with $F=0$ are separated from the states with $|F|=2$. See text for details.}
\end{table*}
%\end{widetext}
%%%%%%%%%%%%%%%%%%%%

Alternatively, one can consider LQ interactions with fermion currents carrying a net $|F|=2$ fermion number
\begin{widetext} 
\begin{align}
\label{eq:yuk-S1}
S_1= \left( \overline{\mathbf{3}}, \mathbf{1},1/3\right):\qquad  \mathcal{L}_{S_1}=& \left( y_{S_1}^L\right)_{ij}\, \overline{Q^C_i} i\tau_2 S_{1} L_{j}+  \left( y_{S_1}^R\right)_{ij}\, \overline{u^C_{R_i}}  S_{1} \ell_{R_j}+\textrm{h.c.}\,,\\[0.5em]
\label{eq:yuk-S3}
S_3= \left( \overline{\mathbf{3}}, \mathbf{3},1/3\right):\qquad  \mathcal{L}_{S_3}=& \left( y_{S_3}^L\right)_{ij}\,  \overline{Q^C_i} i\tau_2 (\vec{\tau} \cdot \vec{S}_{3})  L_{j}+\textrm{h.c.}\,,\\[0.5em]
\label{eq:yuk-S1t}
\widetilde S_1= \left( \overline{\mathbf{3}}, \mathbf{1},4/3\right):\qquad \mathcal{L}_{\widetilde{S}_1}=& \left( y_{\widetilde{S}_1}^R \right)_{ij}\,  \overline{d^C_{R_i} }\widetilde{S}_1 \ell_{R_j} +\textrm{h.c.}\,,
\end{align}
\end{widetext}
where $\Psi^C$ denotes a charge-conjugated fermion. These LQ states are potentially dangerous because they could have diquark couplings which would trigger baryon and lepton number violation~\cite{Dorsner:2016wpm,Assad:2017iib}. Here and after, we assume that these couplings are forbidden by a suitable symmetry that guarantees proton stability.

%%%%%%%%%%%%%%%%%%%%%
\begin{table*}
\renewcommand{\arraystretch}{1.8}
\centering
\begin{tabular}{|c|cc|cc|ccc|}
\hline 
\multicolumn{8}{|c|}{ Neutrinos} \\ \hline\hline
Coupling & $q$ &$l$  &  $R_2$  & $\widetilde{R}_2$ & $S_1$ &  $S_3$  & $\widetilde{S_1}$ \\ \hline\hline
\multirow{2}{*}{$y^L_{ij}$}  & $u_i$ & $\nu_j$ &   $\Big{(}y_{R_2}^L\Big{)}_{ij}$    & $0$    & 	$0$ & $\sqrt{2}\Big{(}V^\ast y_{S_3}^L\Big{)}_{ij}$& $0$ \\%[0.3em] 
         & $d_i$ & $\nu_j$  &  $0$     &  $\Big{(}y_{{\widetilde{R}_2}}^L\Big{)}_{ij}$ & $-\Big{(}y_{S_1}^L\Big{)}_{ij}$ 	& $-\Big{(}y_{S_3}^L\Big{)}_{ij}$  &  $0$\\%[0.3em]
%  \cdashline{1-8}
\hline
\multirow{2}{*}{$y^R_{ij}$} & $u_i$ & $\nu_j$  &  $0$    & $\Big{(}V y_{{\widetilde{R}_2}}^R\Big{)}_{ij}$& 	$0$ & $0$ & $0$ \\%[0.3em]  
&$d_i$ & $\nu_j$  &   $0$    & $\Big{(}y_{{\widetilde{R}_2}}^R\Big{)}_{ij}$ & $\Big{(}y_{S_1}^{\prime\,R}\Big{)}_{ij}$	& $0$ & $0$ \\ 
    \hline
\end{tabular}
\caption{ \sl \small Expressions for the coefficients $y^L_{ij}$ and $y^R_{ij}$ in Eq.~\eqref{eq:yukF0} for each LQ state listed in Sec.~\ref{sec:lq-classification} interacting with neutrinos. See the caption of Tab.~\ref{tab:lq-matching-l}. }
\label{tab:lq-matching-nu} 
\end{table*}
%%%%%%%%%%%%%%%%%%%%

Finally, we provide a prescription to connect the generic couplings $y^L_{ij}$ and $y^R_{ij}$ defined by Eq.~\eqref{eq:yukF0} onto ones of the specific models listed in Eq.~\eqref{eq:yuk-R2}--\eqref{eq:yuk-S1t}. This matching is given in Tabs.~\ref{tab:lq-matching-l} and \ref{tab:lq-matching-nu} for charged leptons and neutrinos, respectively.

\subsection{Higgs and LQ portals to Dark Matter}
\label{sec:lq-dm}

We extend the framework introduced above by adding the DM candidate, a real scalar $S$ not carrying any SM gauge quantum numbers. Stability of $S$ is ensured by a $\mathbb{Z}_2$ symmetry under which $S\to -S$ while other fields remain unchanged. Interactions between dark and visible sectors, at the renormalizable level, proceed via scalar potential couplings. Here, we provide the phenomenological expressions for these interactions which can be viewed as a generalization of the Higgs portal model~\cite{Silveira:1985rk,hep-ph/0011335,1009.5377}.

The general Lagrangian for the scalar sector takes the form
\begin{equation}
  \begin{split}    
    \mc{L}_\mrm{scalars} =&  \left(D_\mu H\right)^\dagger \left( D^\mu H \right) + \left(D_\mu \Delta\right)^\dagger \left( D^\mu \Delta\right) \\ &+ \dfrac{1}{2}\left( \partial_\mu S \right)\left( \partial^\mu S \right) - V(H, \Delta, S) \ ,
      \end{split}
\label{eq:Lscalar}
\end{equation}
where $H$ denotes the SM Higgs doublet field and we consider a single LQ multiplet $\Delta$. Besides the kinetic terms, with covariant derivatives as given in Eq.~\eqref{eq:Dmu}, we have a scalar potential that we parameterize as follows
\begin{equation}
V(H, \Delta, S) = V_\mrm{SM}(H) + V_\mrm{BSM}(H, \Delta, S) \ .
\label{eq:Vfull}
\end{equation}
The first contribution is the same one as in the SM involving just the Higgs doublet
\begin{equation}
V_\mrm{SM}(H) = - \mu^2 |H|^2 + \lambda |H|^4 \ .
\label{eq:VSM}
\end{equation}
Here, we have $\mu^2 > 0$ in order to ensure the spontaneous breaking of the electroweak symmetry. The BSM scalar potential contains masses and interactions of the new degrees of freedom
\begin{equation}
  \begin{split}
V_\mrm{BSM}(H, \Delta&, S) = m_{1}^2 \left|\Delta\right|^2 + \dfrac{m_{2}^2}{2} S^2 +  \dfrac{\lambda_1}{4} S^4 \\ & + \dfrac{\lambda_2}{4} \left| \Delta\right|^4 + 
\dfrac{\lambda_3}{2} S^2 \left|\Delta \right|^2  \\ & +\dfrac{\lambda_4}{2} S^2 \left|H \right|^2+ 
\dfrac{\lambda_5}{2} \left|\Delta \right|^2 \left|H \right|^2  \ ,   
  \end{split}
\label{eq:VBSM}
\end{equation}
where the free real parameters $m_{1,2}$ and $\lambda_i$ are always allowed by the symmetries of the theory. For the case of LQ weak doublets (i.e., for~$\Delta=R_2$ or $\widetilde{R}_2$), the additional term 
\begin{align}
\begin{split}
V_\mrm{BSM}(H, \Delta, S) \supset& \frac{\lambda_6}{2} (H^\dagger \Delta)(\Delta^\dagger H) 
\end{split}
\end{align}
is allowed by electroweak gauge invariance. Once the SM Higgs gets a vacuum expectation value~(vev), this interaction induces a mass splitting among the $SU(2)$ doublet LQ states. This operator, also constrained by $T$-parameter~\cite{Baak:2012kk}, is not relevant for DM phenomenology. Moreover, if more than one scalar LQ is introduced, other operators that mix the different LQs via couplings to the SM Higgs are also allowed~\cite{Hirsch:1996qy}. These operators can be relevant to generate neutrino masses~\cite{Dorsner:2017wwn}, or to generate dipole operators to accommodate the muon $g-2$ anomaly~\cite{Dorsner:2019itg}. Nonetheless, they do not play any relevant role in DM phenomenology. For these reasons, we neglect these further interactions in our study.

Not all couplings in the scalar potential in Eq.~\eqref{eq:VBSM} have the same impact on our analysis. The quartic couplings $\lambda_{1,2}$, which correspond to self-interactions of $S$ and $\Delta$, do not play a considerable role in phenomenological analysis apart from their impact on the stability of the potential. Focusing on interactions that do impact DM phenomenology, we can express the scalar potential after electroweak symmetry breaking as follows
\begin{align}
  \label{eq:VBSM2} 
V_\mrm{BSM}(H, &\Delta, S) \supset m_\Delta^2 |\Delta|^2 + \frac{m_S^2}{2} S^2+ \dfrac{\lambda_3}{2} S^2 \left| \Delta\right|^2   \nonumber
\\ &+ \dfrac{1}{4} (\lambda_4 S^2 + \lambda_5 \left| \Delta\right|^2) (2 v h + h^2) \,.
\end{align}
Here, $h$ is the SM Higgs boson and $v = (\sqrt{2} G_F)^{-1/2}$ is the Higgs vev. DM phenomenology will change accordingly to the values assumed for the quartic couplings $\lambda_{3,4,5}$, and the masses $m_S$ and $m_\Delta$.

\subsection{Stability constraints}

%Stability of the full tree-level scalar potential in Eq.~\eqref{eq:Vfull} requires positivity of the purely quartic terms: $\lambda>0$ and $\lambda_{1,2}>0$. However, in the presence of the other quartic interactions $\lambda_{3,4,5}$ this condition is necessary but not sufficient. 
Here we discuss the theoretical constraints on the quartic couplings $\lambda_i$ arising from the stability of the scalar potential in Eq.~\eqref{eq:Vfull}. The quartic terms in Eqs.~\eqref{eq:VSM} and \eqref{eq:VBSM} can be written as a quadratic form
\begin{equation}
  \label{eq:PotMatrix}
  \begin{split}    
    &V_\lambda(H, \Delta, S)  =\\
    &=\begin{pmatrix} |H|^2 &S^2 &|\Delta|^2 \end{pmatrix}
  \begin{pmatrix}
    \lambda & \lambda_4/4 & \lambda_5/4 \\
    \lambda_4/4 & \lambda_1/4 & \lambda_3/4 \\
    \lambda_5/4 & \lambda_3/4 & \lambda_2/4
  \end{pmatrix}
  \begin{pmatrix} |H|^2 \\ S^2 \\ |\Delta|^2 \end{pmatrix}.
    \end{split}
\end{equation}
The above expression has to be bounded from below for large field values in any arbitrary direction. If we consider going to infinity along the direction of a single field, $|H|^2 \to \infty$, $S^2 \to \infty$, or $|\Delta|^2 \to \infty$, this results in necessary conditions $\lambda,\lambda_1,\lambda_2 > 0$, respectively.

In order to find the sufficient conditions, we diagonalize the $n\times n$ matrix appearing in Eq.~\eqref{eq:PotMatrix} and require all its eigenvalues $x_{1,2,\ldots,n}$ to be positive. This quadratic form is diagonalizable and thus its characteristic polynomial $p(x)$ has $n$ roots $x_{1,\ldots,n}$ and $n-1$ stationary points $e_{1,\ldots,n-1}$ which are ordered as
\begin{equation}
  x_1 \leq e_1 \leq x_2 \leq \ldots \leq x_{n-1} \leq e_{n-1} \leq x_n.
\end{equation}
If some of the roots are multiple, then the above inequalities become equalities across the range spanned by multiple roots (e.g., for a double root $x_1 = x_2$ we have $x_1 = e_1 = x_2$).

\begin{figure}
  \centering
  \includegraphics[scale=0.8]{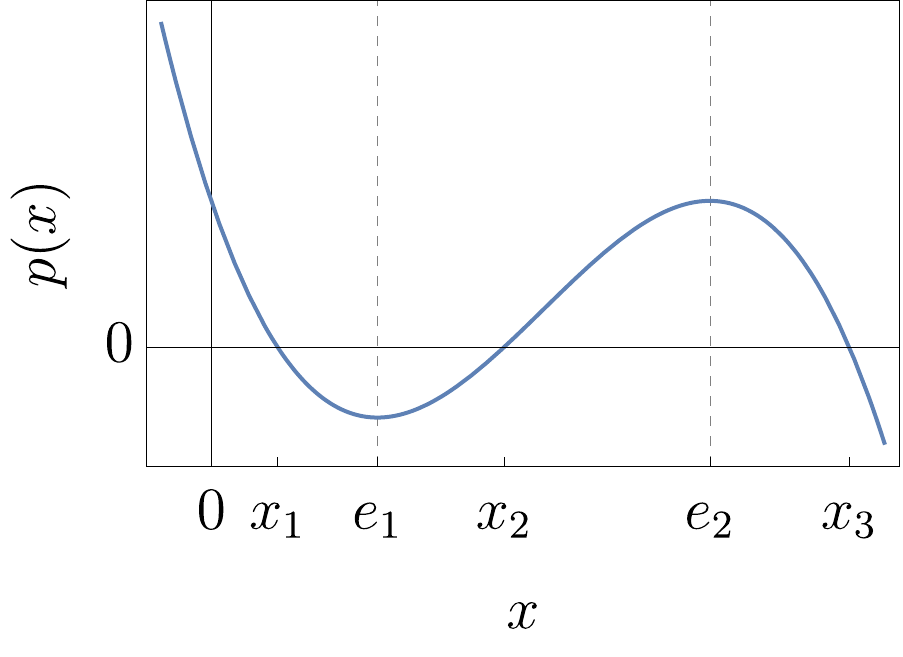}
  \caption{Characteristic polynomial $p(x)$ when all its roots are positive.}
  \label{fig:charpoly}
\end{figure}

We show in Fig.~\ref{fig:charpoly} the characteristic polynomial $p(x)$ for the case when all roots are positive. As we can see, the requirement $x_1 \geq 0$ is fulfilled precisely when $e_1 \geq 0$ and $p(0) \geq 0$ since $p(x) > 0$ for large negative $x$. If we repeat the above argument for $e_1 \geq 0$, it follows that $p'(0) \leq 0$ and that the first stationary point of $p'(x)$ should be non-negative. Iterating this argument to yet higher derivatives yields the conditions\footnote{This is a special case of the so-called \emph{Descartes' rule of signs}.}
\begin{align}
  p(0) \geq 0,&\quad p'(0) \leq 0\,,\quad p''(0) \geq 0\,,\quad \ldots,\nonumber \\ & \ldots, \quad p^{(n)}(0) (-1)^n \geq 0\,.
\end{align}
Thus sufficient stability conditions can be expressed in terms of $\lambda_i$. We find it convenient to define the parameters 
\begin{align}
\label{eq:tetrahedron2}
x \equiv \frac{\lambda_4}{2\sqrt{\lambda \lambda_1}}\,, \quad y \equiv \frac{\lambda_5}{2\sqrt{\lambda \lambda_2}}\,, \quad z  \equiv \frac{\lambda_3}{\sqrt{\lambda_1 \lambda_2}}\,.
\end{align}
In addition to the already mentioned positivity of $\lambda$ and $\lambda_{1,2}$, we derive the following sufficient stability conditions
  \begin{align}
    \label{eq:tetrahedron}
    \begin{split}
  |x|, |y|, |z| &< 1\,, \\[0.4em]
    x^2 + y^2 + z^2 -2xyz &< 1\,.
    \end{split}
  \end{align}
  The inequalities~\eqref{eq:tetrahedron} define a tetrahedron-like
  region in the $\lbrace x, y, z\rbrace$ space. 
  
  Are the above
  conditions also necessary? It turns out that since the potential
  matrix is multiplied by strictly positive vectors from the left and
  right, the above condition can be
  somewhat relaxed. We require that matrix $V$ is co-positive
  definite, which by definition means that $x^T V x > 0$ for any vector $x$ from the first
  octant.  Following Ref.~\cite{1205.3781}, in particular Eqs.~(5) and
  (6) of that paper for particular case of $3\times 3$ matrices, we find slightly looser conditions compared to~\eqref{eq:tetrahedron}:
    \begin{align}
	    \label{eq:tetrahedronCop}
  x, y, z &>  -1\,, \\[0.4em]
    x^2 + y^2 + z^2 -2xyz &< 1\, \quad \text{for }x+y+z < -1\,. \nonumber
    \end{align}
    Together with $\lambda, \lambda_1, \lambda_2 > 0$, these are necessary and sufficient stability conditions.
   The most interesting among these inequalities are the ones for $\lambda_{3,4,5}$ since they are related to the interactions between the scalar particles in our model. In particular, we can derive the following necessary conditions on $\lambda_{3,4,5}$,
  \begin{align}
      \label{eq:finalbounds}
      \begin{split}
     -3.5 \lesssim   \lambda_3 &\lesssim  3.5\,,\\[0.3em]
       -1.4 \lesssim  \lambda_4 &  \lesssim  3.5\,,\\[0.3em]
      -1.4 \lesssim  \lambda_5 & \lesssim  3.5\,,
    \end{split}
 \end{align}
where the leftmost bounds are realised when both $\lambda_1$ and $\lambda_2$ are set to the perturbativity limit $\lambda_{1,2} = \sqrt{4\pi}$ and inserted to the stability lower bounds. The rightmost upper bounds stem from the perturbativity requirement, whereas stability considerations do not yield an upper bound on $\lambda_{3,4,5}$.

% perturbative unitarity: further constraints on the quartic couplings can be derived from...

%%%%%%%%%%%%%%%%%%%%%%
%%%%%%%%%%%%%%%%%%%%%%
%%%%%%%%%%%%%%%%%%%%%%
\section{Higgs and Leptoquark Portals Phenomenology}
\label{sec:phenomenology}
%%%%%%%%%%%%%%%%%%%%%%
%%%%%%%%%%%%%%%%%%%%%%
%%%%%%%%%%%%%%%%%%%%%%

In this section, we study the phenomenology of our scalar leptoquark portal framework. DM phenomenology is driven by the scalar potential interactions in Eq.~\eqref{eq:VBSM2}, and the scalar singlet $S$ interacts with the visible world via the Higgs portal ($\lambda_4$ coupling) as well as through the LQ portal ($\lambda_3$ coupling). In this section, we  compute the DM relic density, we study constraints from direct and indirect searches and we account for collider bounds. 

\subsection{Relic density}

The DM number density $n_S$ evolves according to the Boltzmann equation~\cite{Gondolo:1990dk,Steigman:2012nb}
%%%%%%%%%%%%%%%%
\begin{equation}
\label{eq:boltzmann_n}
\frac{\mathrm{d}n_S}{\mathrm{d}t} + 3 H n_S = - \braket{\sigma v} \left(n_S^2 - n_S^{\mathrm{eq} \, 2}\right) \ .
\end{equation}
%%%%%%%%%%%%%%%%
Here, the Hubble parameter $H = \frac{1}{a} \frac{da}{dt}$ accounts for dilution due to the expansion and it is defined in terms of the cosmological scale factor $a$ and its derivative with respect to the cosmic time $t$. The change in the number of $S$ particles due to collisions is accounted for by the term on right-hand side, proportional to the thermally averaged annihilation cross section times the M{\o}ller velocity $\braket{\sigma v}$. The inverse process, where degrees of freedom from the thermal bath annihilate to produce a pair of DM particles, is proportional to the square of the DM equilibrium number density $n_S^{\mathrm{eq}}$.

When DM particles are in equilibrium in the early universe, their energies and momenta are distributed in phase space accordingly. Thus the kinematics of the initial state for each DM annihilation is not fixed, and we need to account for all possible options. This is the reason why we have a thermal average in Eq.~\eqref{eq:boltzmann_n}. There is a general expression
\begin{widetext}
\begin{equation}
\langle \sigma v\rangle_{S S \rightarrow ij} = \frac{1}{8 \, m_S^4 \, T \, K_2^2[m_S / T]} 
\int_{4m_S^2}^{\infty} ds \,( s -4 m_S^2) \, \sqrt{s} \, \sigma_{S S \rightarrow ij}(s) \,K_1\left(\frac{\sqrt{s}}{T}\right) \ , 
\label{eq:thermal-average}
\end{equation}
\end{widetext}
connecting the annihilation cross section as a function of the Mandelstam variable $s$, which is nothing but the (square of the) center of mass energy, and the thermally averaged cross section. Here, $T$ is the temperature of the primordial thermal bath and $K_{1,2}(x)$ are modified Bessel functions of the second kind. This expression is valid for a Maxwell-Boltzmann statistics of initial state DM particles, and we can trust it since quantum degeneracy effects give very small corrections in the early universe. In order to determine the relic density, we need to compute expressions for all the DM annihilations allowed by kinematics. 

The phenomenologically relevant DM annihilation channels to visible final states are: (i) $SS \to hh$, (ii) $SS\to \Delta\Delta$, (iii) $SS\to VV$ (gauge bosons) and (iv) $SS\to ff$ (fermions). We account for the leading contributions for each channel, which can appear either at tree or one-loop level and depend on the underlying parameters of the scalar potential as well as leptoquark flavour parameters. These expressions should be convoluted with the thermal distributions at finite temperature $T$ as prescribed by Eq.~\eqref{eq:thermal-average}. We provide full derivations and complete expressions for the annihilation cross section in App.~\ref{app:AnnihilationXSec} and we evaluate the DM relic density by following standard techniques~\cite{Gondolo:1990dk}.

The leading annihilation channel, among the several ones available, depends on what parameter space region we focus on. For illustration, we consider two different benchmarks where we fix the quartic couplings $\lambda_{3,4,5}$. We neglect for this discussion the LQ Yukawa couplings to SM fermions; we comment their impact on DM phenomenology in Sec.~\ref{sec:results}.

\begin{description}

\begin{figure*}
  \centering
  \includegraphics[width=.3\linewidth]{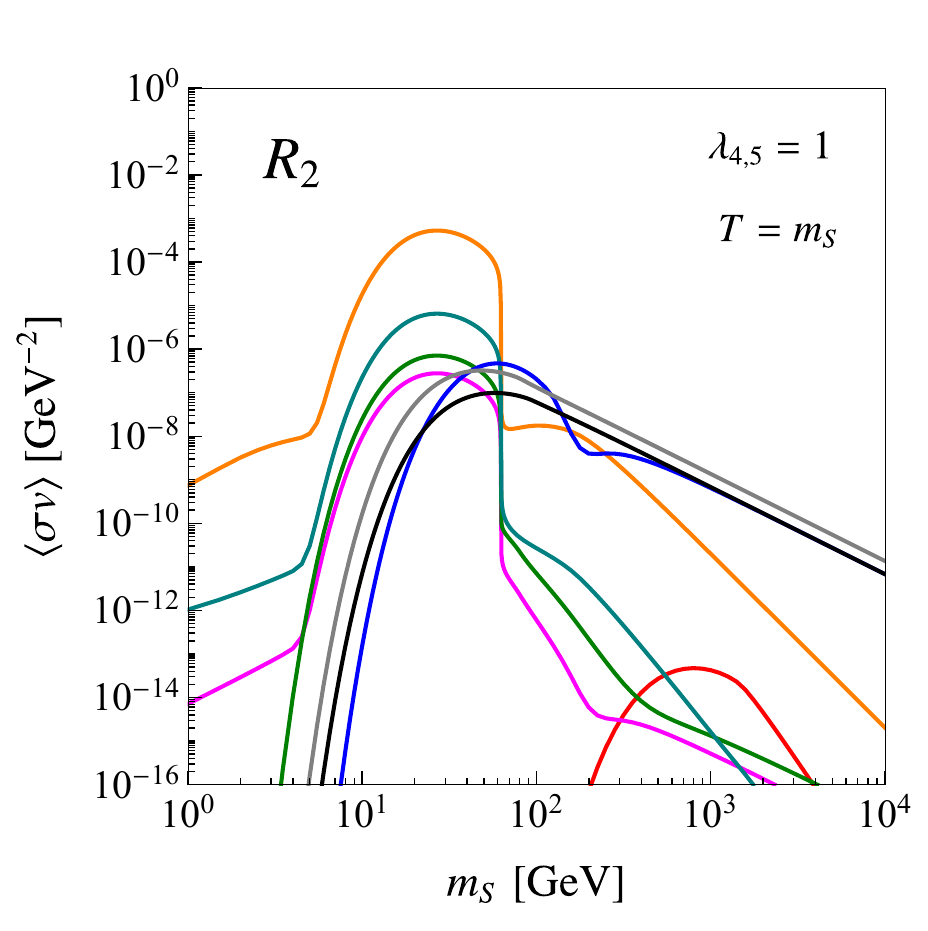}
  \includegraphics[width=.3\linewidth]{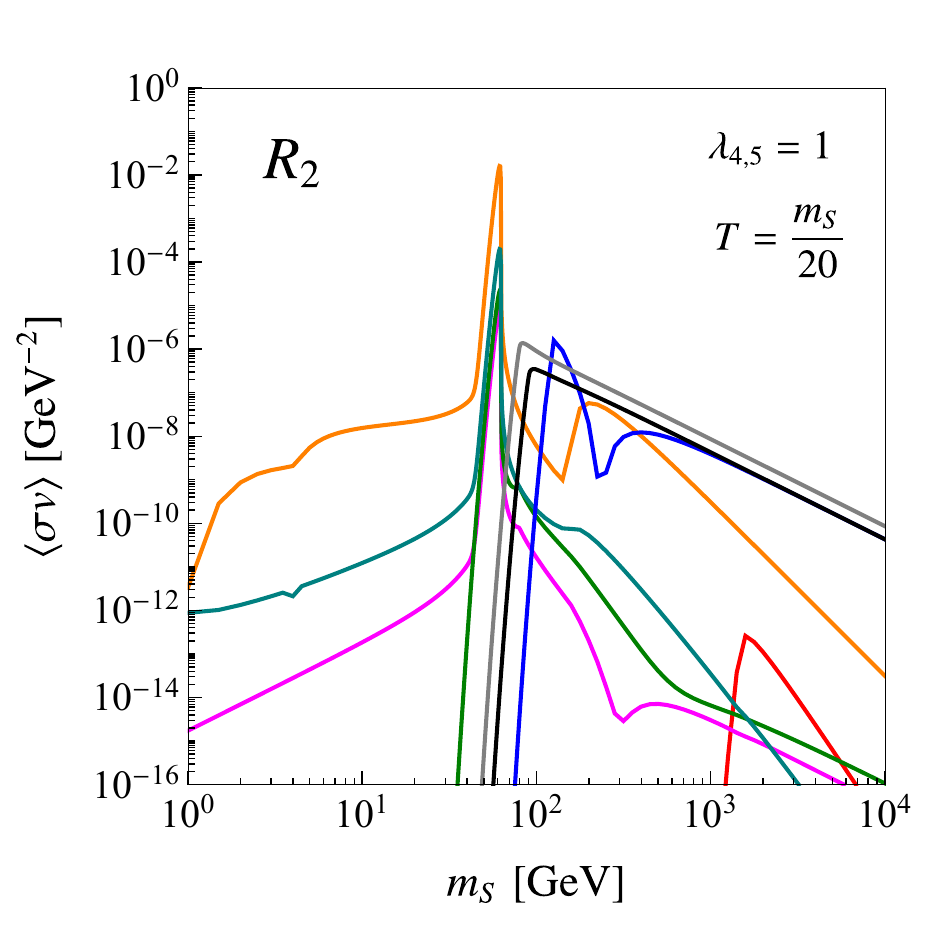}
  \includegraphics[width=.365\linewidth]{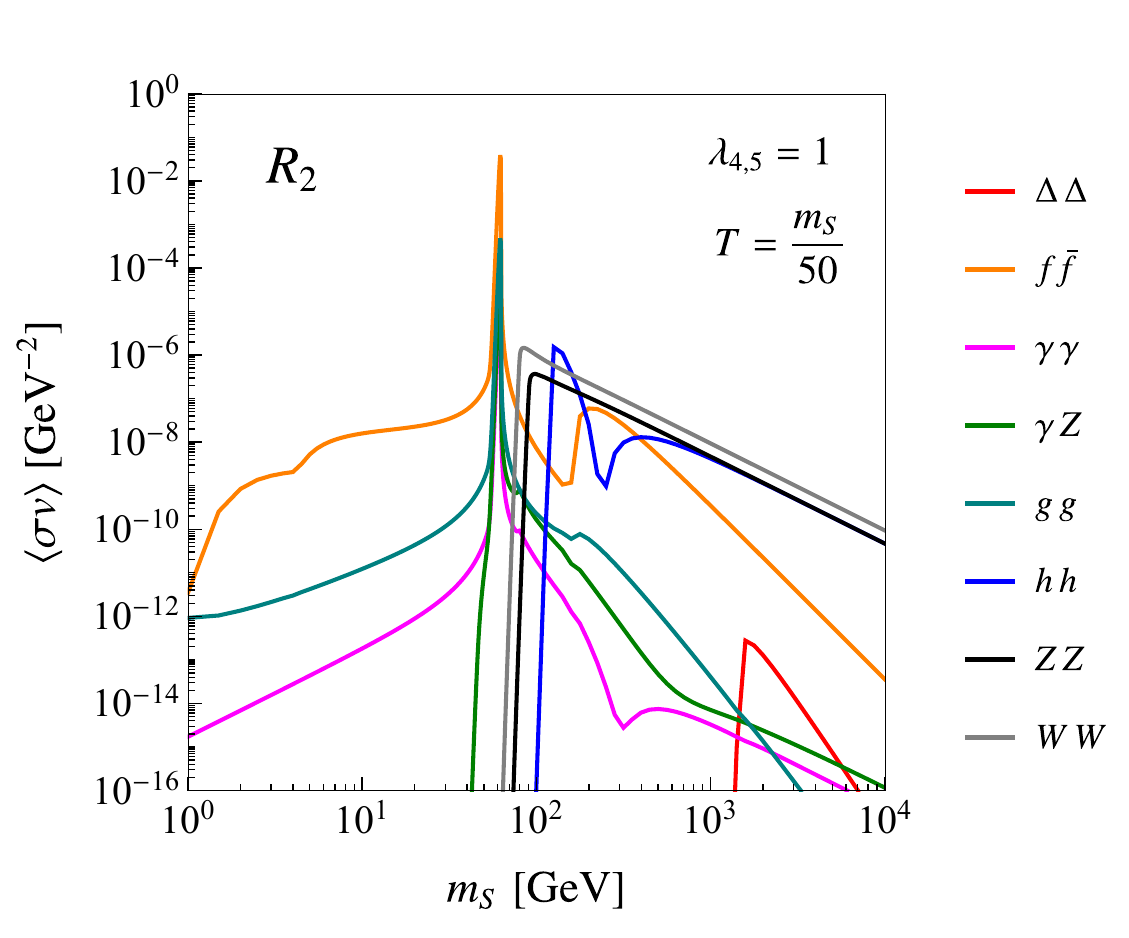}
  \caption{\small \sl Thermally-averaged annihilation cross sections as functions of the DM mass $m_S$ for the Higgs portal benchmark ($\lambda_3 = 0$, $\lambda_{4,5}=1$). We consider the three temperature values around the epoch of freeze-out: (i) $T=m_S$ (left panel), (ii) $T=m_S/20$ (right panel) and (iii) $T=m_S/50$ (bottom panel). For illustration, we choose the LQ state $R_2 = (\bm{3},\bm{2},7/6)$ with mass $1.5\e{TeV}$. However, the main features remain similar when considering other LQ states.}
  \label{fig:T-avg}
\end{figure*}

\begin{figure*}
  \centering
  \includegraphics[width=.3\linewidth]{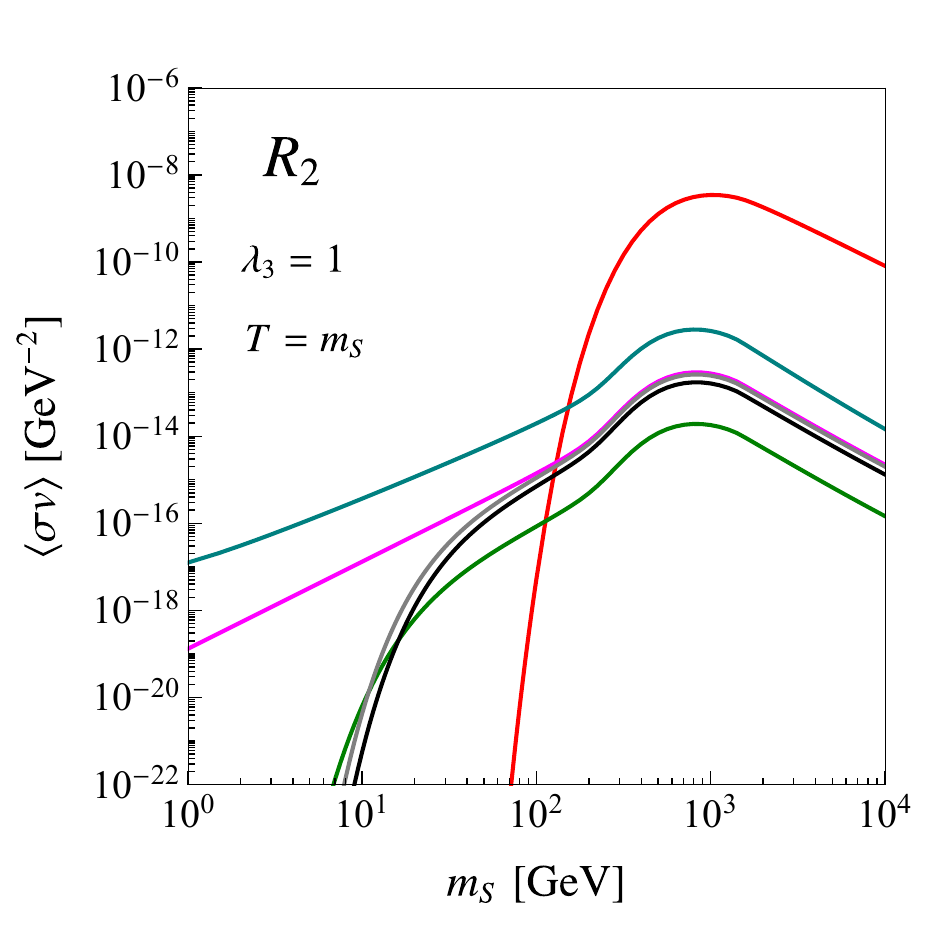}
  \includegraphics[width=.3\linewidth]{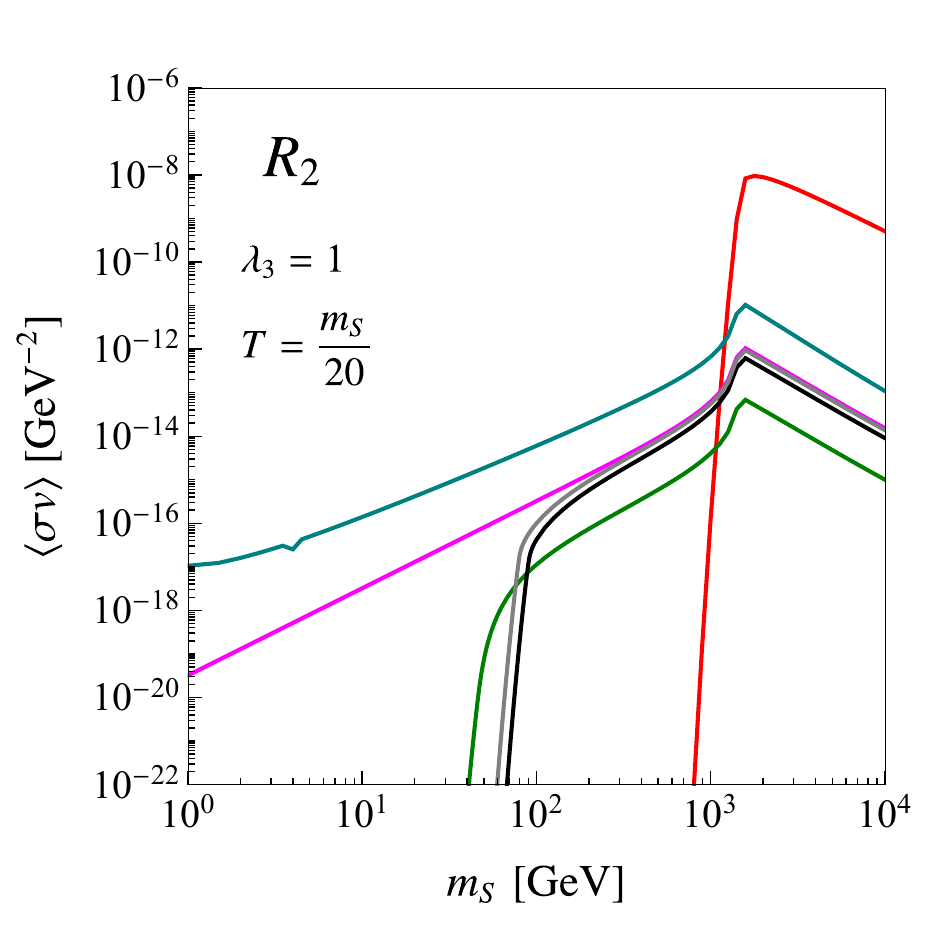}
  \includegraphics[width=.365\linewidth]{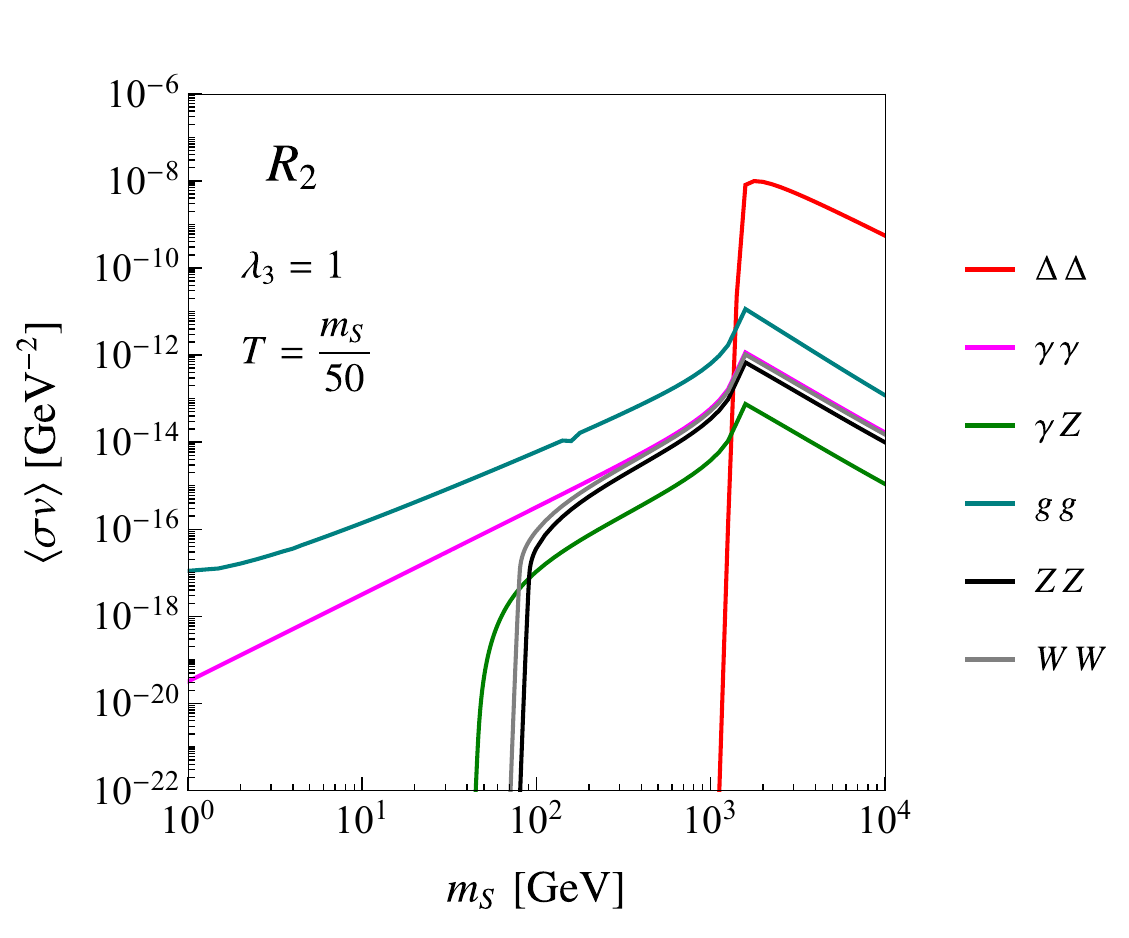}
  \caption{\small \sl Same as Fig.~\ref{fig:T-avg} but for the LQ portal benchmark ($\lambda_{3}=1$, $\lambda_{4, 5} = 0$).}
  \label{fig:T-avg-2}
\end{figure*}

\item[Higgs portal: $\lambda_3=0$, $\lambda_{4, 5} = 1$] The leading annihilation channels are driven by the $\lambda_4$ quartic coupling with the SM Higgs field. The $\lambda_5$ quartic interaction between LQ and Higgs field plays only a marginal role since it only contributes to the subdominant channel $SS \to h \to \Delta \Delta^*$ proportionally to $\lambda_4\, \lambda_5$. We show the thermally-averaged annihilation cross sections as functions of the DM mass $m_S$ for this scenario in Fig.~\ref{fig:T-avg}. For the purpose of illustration, the LQ state $\Delta$ is in the $R_2 = (\mathbf{3},\mathbf{2},7/6)$ representation of the SM gauge group. We provide three snapshots at three different temperatures $T/m_S \in \lbrace 1, 1/20, 1/50 \rbrace$; this is the relevant temperature range for thermal freeze-out of cold relics. Annihilations via $s$-channel Higgs exchange are by far dominant. The dominant final states can be either $f \bar f$ or $V_1 V_2$ (with $V_{1,2}$ a SM electroweak gauge boson), depending on the mass of $S$, whereas annihilations to LQs are sub-dominant since LQs do not couple directly to DM in this scenario. For this reason, LQ interactions only contribute indirectly to the relic density via the modification of the $h V_1 V_2$ couplings, which are mostly relevant for $m_S$ values around the electroweak scale. The thermally averaged cross sections have a peak at the Higgs pole, $m_S = m_h/2$, which becomes wider but still dominant as the temperature $T$ increases. We comment about the (in)viability of this scenario to account for the observed DM relic density in Sec.~\ref{sec:results}.

\item[LQ portal: $\lambda_3=1$, $\lambda_{4, 5} = 0$] In this case, the situation is inverted since DM does not couple at tree-level to the Higgs boson whereas it has $\mathcal{O}(1)$ couplings with LQs. Once we set $\lambda_{4} = 0$, this scenario is actually completely blind to $\lambda_5$ since it turns out that only a combination of $\lambda_4 \lambda_5$ enters the annihilation cross sections. From Fig.~\ref{fig:T-avg-2}, we see that the dominant annihilation channel is to LQ final states, which become kinematically accessible above the threshold $m_S > m_\Delta$, via the $\lambda_3$ quartic interaction. For smaller values of $m_S$, the dominant modes are $SS\to V_1 V_2$ induced by LQ loops. Cross sections in Fig.~\ref{fig:T-avg-2} are much smaller than the ones appearing in Fig.~\ref{fig:T-avg} since they are loop suppressed, except for the $\Delta \Delta^*$ channel, and since they are not resonantly enhanced as in the case of the Higgs portal regime. Here, the relevant parameter space will appear at large DM masses where direct detection constraints are less effective, as we discuss in Sec.~\ref{sec:results}.

\end{description}

The above discussion was based on the $R_2$ representation for the LQ state. Considering different representations would only affect the annihilation channels into the gauge bosons $V_1 V_2$, since they depend on the $SU(2)_L$ and $U(1)_Y$ quantum numbers of the LQs running in the loops, see e.g.~Eq.~\eqref{eq:DVVdirect} in App.~\ref{app:loop}, which amount to mild modifications of the annihilation cross section. On the other hand, the $\Delta \Delta^*$ channel is enhanced for LQs with larger multiplicity $2T+1$, but without changing the general features described above. 

\subsection{Direct detection}
\label{ssec:direct-detection}

\begin{figure*}[!t]
\centering	
\includegraphics[width=.47\linewidth]{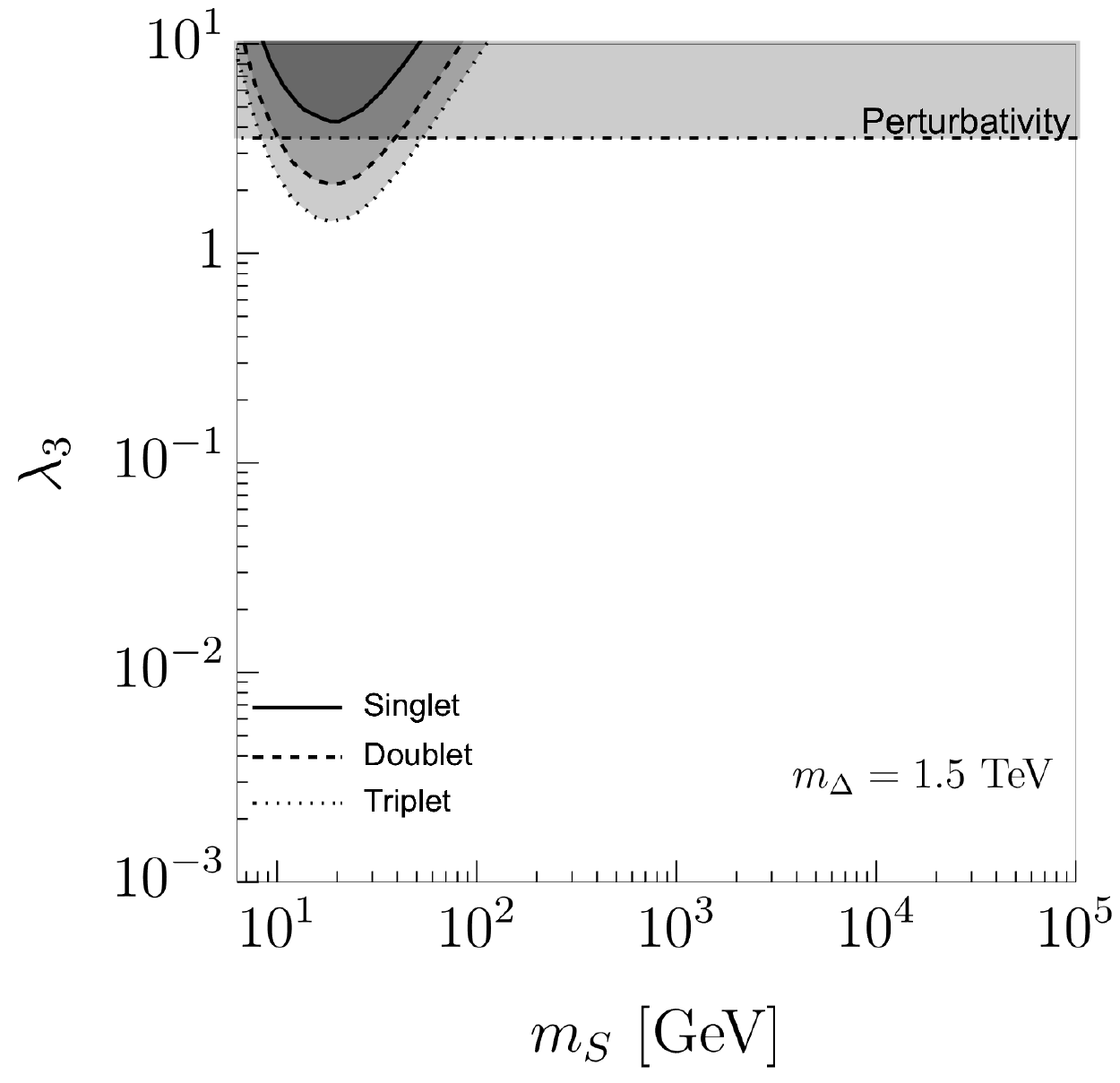} $\quad$
\includegraphics[width=.47\linewidth]{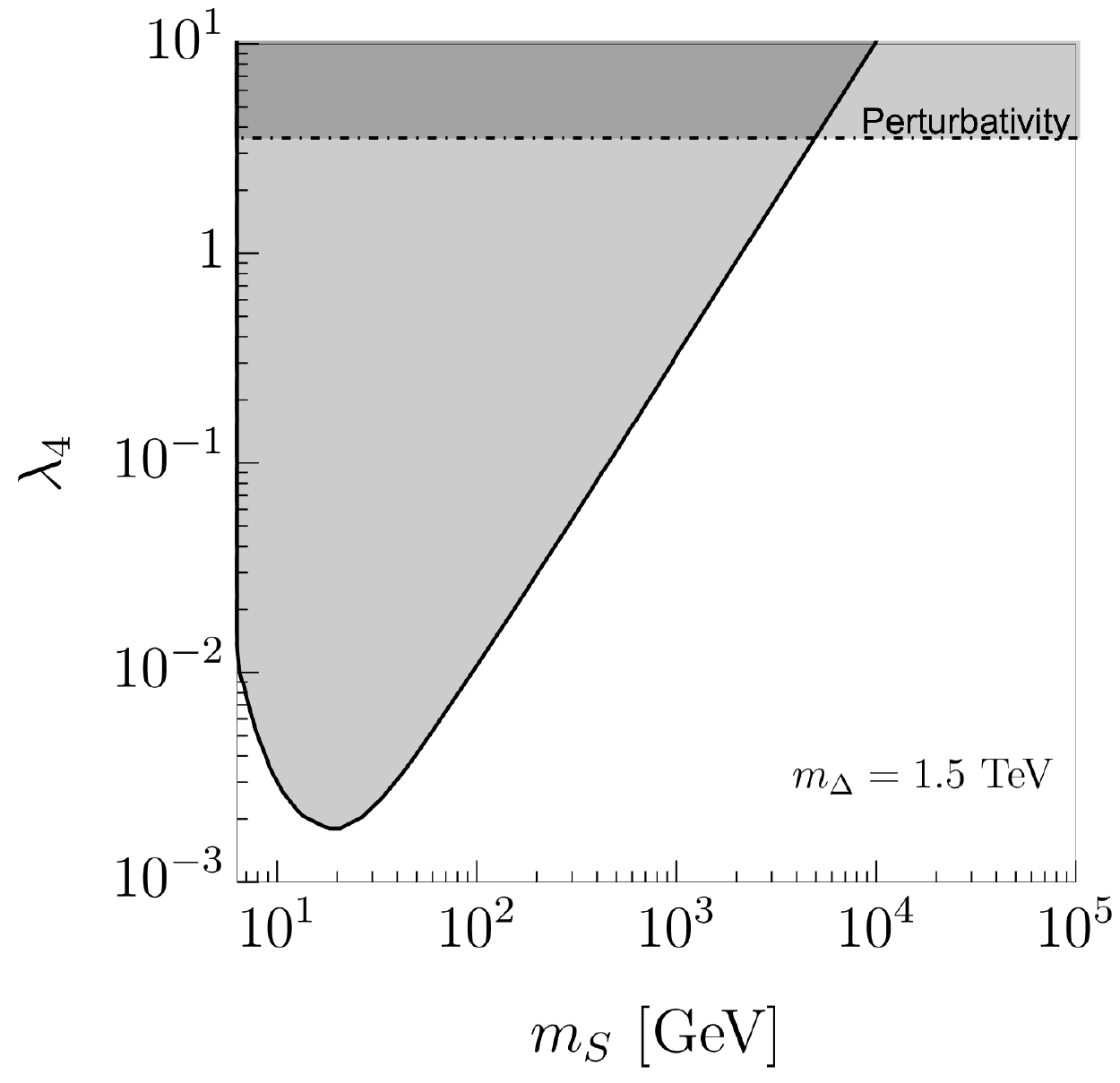} 
\caption{\small \sl Parameter space regions excluded by XENON1T~\cite{Aprile:2018dbl} in the plane $m_S$ vs.~$|\lambda_3|$ (left panel) and $m_S$ vs.~$|\lambda_4|$ (right panel). We fix $m_\Delta=1.5$~TeV and the other couplings are set to zero. Constraints on $\lambda_3$ depend on the specific $SU(2)_L$ representation (singlet, doublet or triplet), while the ones on $\lambda_4$ are, to first approximation, independent of LQ interactions. The coupling $\lambda_5$ cannot affect direct detection constraints (see text).}
\label{fig:direct-detection}
\end{figure*}

DM effective interactions with nucleons $N$ mediate elastic scattering that can be searched for by direct detection experiments. The effective operator relevant to our analysis is
%%%%%%%%%%%%%%%%
\begin{equation}
\mathcal{L}_{N}^{\mathrm{eff}} = C_N \, S^2 \bar{N}N \,,
\end{equation}
%%%%%%%%%%%%%%%
where $C_N$ is a low-energy Wilson coefficient. From the microscopic point of view, these interactions arise from DM couplings to quarks and gluons. The corresponding Lagrangian, defined at the scale $\mu = \mathcal{O}(1~\mathrm{GeV})$ above confinement, is~\cite{DelNobile:2013sia}
%%%%%%%%%%%%%%%%
\begin{equation}
\label{eq:leff-dd}
\mathcal{L}_{\rm QCD}^{\mathrm{eff}}= \sum_{q=u,d,s} C_q \, S^2 \, \bar{q}q+ C_g\, S^2\, \dfrac{\alpha_2}{12\pi} G_{\mu\nu}^a G^{\mu\nu\,a} \ .
\end{equation}
Only effective couplings to light quarks, $C_q$, appear in the Lagrangian since heavy quarks ($c,b,t$) have been integrated out. Their virtual effects provide further contributions, besides the ones due to new physics, to the effective couplings to gluons $C_g$. We can match these effective coefficients to quarks and gluons onto the nucleon Effective Field Theory~\cite{Shifman:1978zn}
%%%%%%%%%%%%%%%%
\begin{equation}
C_N = \sum_{q=u,d,s} \dfrac{C_q}{m_q} \, m_N\, f_{Tq}^{N}- C_g \, \dfrac{2}{27}\, m_N\,f_{TG}^N\,,
\end{equation}
The quantities $f_{Tq}^N$ and $f_{Tg}^N$ are set by nucleonic matrix elements. We use the following values: $f^p_{T_u}=0.023$, $f^p_{T_d}=0.032$ and $f^p_{T_s}=0.020$ for a proton and $f^n_{T_u}=0.017$, $f^n_{T_d}=0.041$ and $f^n_{T_s}=0.020$ for neutrons, with $f^{p,n}_{TG}=1-\sum_{q=u,d,s} f^{p,n}_{Tq}$ \cite{Hisano:2010yh}. This operator induces a spin-independent scattering between the DM particle and the target nucleus. We report the cross section for this scattering normalized per nucleon 
\begin{equation}
\sigma_{\mathrm{SI}} = \dfrac{\mu_{p}^2}{\pi m_S^2}\Big[\frac{Z\, C_p + (A-Z)\, C_n}{A}\Big]^2\,,
\end{equation}
where $Z$ and $A$ are the atomic and mass numbers of the nucleus, and $\mu_p= m_p m_S/(m_p+m_S)$ is the reduced mass for the DM-proton system.

The effective low-energy couplings for our LQ scenario can be read from Eq.~\eqref{eq:Dgg}
\begin{align}
\label{eq:C1q}
C_q &= \dfrac{\lambda_4 m_q}{2 m_h^2}\,,\\
\label{eq:C1g}
C_g &= \dfrac{\lambda_3}{8 m_\Delta^2}(2T+1)-\dfrac{\lambda_4}{2 m_h^2}\left[n_H+\dfrac{\lambda_5 (2T+1)v^2}{16 m_\Delta^2}\right]\,,
\end{align}
%%%%%%%%%%%%%%%%
where $q \in \lbrace u,d,s\rbrace$ and $n_H=3$ is the number of heavy quarks ($H$). Contributions of order $1/m_H$ in the heavy-quark expansion have been neglected. These expressions extend results already existing in the literature~\cite{Choi:2018stw}, and the new contributions we compute give order one corrections to the scattering cross section. In particular, the contribution proportional to $\lambda_5$ gives a small correction due to the additional suppression factor $v^2/m_\Delta^2$. 

The current most stringent bounds on spin-independent DM-nucleon cross section come from the XENON1T experiment \cite{Aprile:2018dbl}. In Fig.~\ref{fig:direct-detection}, we show the constraints on $|\lambda_3|$ (left panel) and $|\lambda_4|$ (right panel) as a function of $m_S$, for a benchmark value $m_\Delta=1.5$~TeV. The most constrained coupling turns out to be $\lambda_4$, independently of the LQ mass, and we find that the dominant contribution comes from the Higgs gluonic penguins.

\subsection{Indirect detection}
\label{ssec:indirect-detection}

DM annihilations produce final state photons that are searched for with gamma-ray telescopes. The galactic center (GC) is a natural target and it comes with pros and cons. On one hand, it is not excessively far away and we do not lose too much flux. On the other hand, it is a very active region and astrophysical backgrounds are significant and not completely understood. Dwarf spheroidal satellite galaxies~(dSphs) of the Milky Way are immune from this issue since they are DM-dominated and with considerably fewer backgrounds. In our study, we consider bounds from both of these sources. The strongest constraints for DM masses around or below the weak scale come from the Fermi Large Area Telescope (Fermi-LAT)~\cite{1506.00013,1503.02641,1611.03184}. Different instruments provide meaningful bounds on TeV scale DM candidates: the High Energy Stereoscopic System~(HESS) observatory~\cite{1607.08142,1805.05741,2008.00688}, the Very Energetic Radiation Imaging Telescope Array System (VERITAS)~\cite{Archambault:2017wyh}, and the Major Atmospheric Gamma Imaging Cherenkov Telescopes (MAGIC)~\cite{Ahnen:2017pqx}. For even larger masses, bounds from the High Altitude Water Cherenkov (HAWC) Observatory~\cite{Abeysekara:2017mjj} are the most severe ones. Besides imposing current experimental constraints, we also exploit the discovery reach of the future Cherenkov Telescope Array (CTA)~\cite{Acharyya:2020sbj}.

Experimental collaborations provide bounds on the annihilation cross section as a function of the DM mass for fixed SM final states. There are several annihilation channels available in our framework, and we identify the dominants ones for the two benchmarks introduced at the beginning of this Section: the Higgs portal and the LQ portal. Higgs and weak gauge boson final states are the dominant channels for the former case. Performing an analysis by computing the gamma-ray spectrum from all different final states is beyond the scope of our work, and we provide conservative bounds by imposing the most severe constraints among the ones for the final state particles; our indirect detection bounds are thus rather conservative for this benchmark. For the LQ portal case, we have two LQs in the final state that decay subsequently to quarks and leptons; the DM annihilate to four final state SM particles~\cite{Elor:2015tva}. We compute the expected photon spectrum from annihilations to LQs and we compare with the experimental sensitivities of current and future experiments. As we discuss in the next Section, the region of our interest is the one where the DM mass is above the weak scale. We can establish the dominant annihilation channels for our benchmarks by looking at the right panel of Figs.~\ref{fig:T-avg} and \ref{fig:T-avg-2}; DM particles are non-relativistic today and therefore annihilations are correctly described by thermal averages at low temperatures. 

\subsection{Collider constraints}
\label{sec:collider-constraints}

We conclude this phenomenological section with a list of collider constraints. Conventional mono-X searches for DM do not provide the most stringent bounds from collider physics for our framework. The scalar potential couplings of the Higgs doublet $H$ to $S$ and $\Delta$ leave an imprint on Higgs properties measured at the LHC. Thus we consider Higgs physics as well as direct searches for LQs.

\subsubsection{Invisible Higgs Decay} For a DM particle coupled to the Higgs boson which is light, $m_S < m_h/2$, an important constraint comes from the upper bound on the branching fraction of Higgs decays into invisible final states~\cite{Burgess:2000yq,DEramo:2007anh,Andreas:2010dz,Kanemura:2010sh,Mambrini:2011ik,Djouadi:2011aa}. Within our framework, we have the requirement
\begin{equation}
  \label{eq:hinv-constr}
  \frac{ \Gamma_{h \to SS}}{\Gamma_h^\mrm{SM} +  \Gamma_{h \to SS} } < \mrm{Br}(h \to \mathrm{inv.})  \ ,
\end{equation}
where $\Gamma_h^\mrm{SM} = 4.1\e{MeV}$~\cite{1610.07922}. The invisible partial decay width results in
\begin{equation}
  \label{eq:Gammah-SS}
  \Gamma_{h \to SS}  = \frac{\lambda_4^2 v^2}{32 \pi m_h} \sqrt{1-\frac{4m_S^2}{m_h^2}} \ .
\end{equation}
We impose the experimental bound $\mrm{Br}(h \to \mathrm{inv.}) < 0.24$~\cite{Tanabashi:2018oca} and we find $|\lambda_4| \lesssim 0.02$ for $m_S$ values below $50\e{GeV}$. The constraint is significantly relaxed when $m_S$ is just below $m_h/2$, according to phase space suppression in Eq.~\eqref{eq:Gammah-SS}. This constraint excludes the viability of the low-mass DM since the annihilation cross section of $SS$, driven by the Higgs portal, $\lambda_4$, is too small which leads to overabundance of $S$.

\subsubsection{Higgs Decay to Leptons} The measurements of $h \to \mu^+ \mu^-$ and $h \to \tau^+ \tau^-$ provide important constraints on the $\mu t$ and $\tau t$ leptoquark Yukawas via the modification of the Higgs Yukawa in Eq.~\eqref{eq:ModYuk}. Recently, the CMS collaboration\footnote{See also the recent ATLAS results on the same observable, $\mu_{\mu^+\mu^-} = 1.2 \pm 0.6$~\cite{Aad:2020xfq}.} provided the best-fit values of the signal strength parameters as~\cite{CMS:2020eni}
  \begin{equation}
    \begin{split}     
      \mu_{\mu^+\mu^-} &= \dfrac{\sigma(pp\to h)\, \mathcal{B}(h\to\mu\mu)}{\left[\sigma(pp\to h)\, \mathcal{B}(h\to\mu\mu)\right]_\mathrm{SM}}\\ &= 1.19^{+0.41}_{-0.39}{}^{+0.17}_{-0.16}\,.
          \end{split}
\end{equation}
For the $\tau\tau$ channel, the CMS collaboration also provided~\cite{CMS:2020dvp} 
\begin{equation}
  \begin{split}
    \mu_{\tau^+\tau^-} &= \dfrac{\sigma(pp\to h)\, \mathcal{B}(h\to\tau\tau)}{\left[\sigma(pp\to h)\, \mathcal{B}(h\to\tau\tau)\right]_\mathrm{SM}}\\
    &= 0.85^{+0.12}_{-0.11}\,.        
  \end{split}
 \end{equation}
 Using these constraints and assuming that the modification on the Higgs production due to LQs is negligible, we obtain the following $1\sigma$ constraints from Eq.~\eqref{eq:ModYuk} of App.~\ref{app:AnnihilationXSec} on products of the LQ Yukawa couplings, 
\begin{align}
  \sqrt{|y_L^{t\mu}y_R^{t\mu}|} &< 0.6\times \bigg{(}\frac{m_\Delta}{1~\mathrm{TeV}} \bigg{)}\,,\\
  \sqrt{|y_L^{t\tau}y_R^{t\tau}|} &< 2.3\times \bigg{(}\frac{m_\Delta}{1~\mathrm{TeV}} \bigg{)}\,.
\end{align}

\subsubsection{Higgs production at the LHC} The production cross section of the Higgs boson at the LHC via $gg \to h$ and its decay $h \to \gamma\gamma$ are affected by the virtual LQ loops, proportional to the $\lambda_5$ coupling. In the approximation where these are the dominant effects of leptoquark loops,
we can use the combined result of ATLAS and CMS~(Fig.~17 of Ref.~\cite{1606.02266}), which constrains the relative SM coupling modifiers $\kappa_g$ and $\kappa_\gamma$. Explicit expressions for $\kappa_{g,\gamma}$ are
\begin{align}
  \label{eq:kappa-g}
    \kappa_\gamma &=  1 + \frac{\lambda_5 v^2}{4 m_{\Delta}^2}  N_c (2T+1)[Y^2+T(T+1)/3]\times \nn \\ &\qquad\qquad\times\frac{\mathcal{A}_0(x_{\Delta})}{\mc{A}_1(x_W)+\frac{4}{3}\mc{A}_{1/2}(x_t)}\,,\\
    \kappa_g &=  1 + \frac{\lambda_5  v^2}{4m_\Delta^2}  (2T+1) \frac{\mc{A}_0(x_\Delta)}{\mc{A}_{1/2}(x_t)}\,,
\end{align}
where $x_i = m_h^2/(4m_i^2)$, as in Eqs.~\eqref{eq:higgs-photon} and~\eqref{eq:higgs-gg} in App.~\ref{app:loop}. We find that the current experimental precision on $\kappa_{g,\gamma}-1 \lesssim 0.2 \approx \lambda_5 {v^2}/{(4 m_\Delta^2)}$ results in rather weak constraint $\lambda_5/(4m_\Delta^2) \lesssim 3/\mrm{TeV}^2$.

%%%%%%%%%%%%%%%%%%%
%%%%%%%%%%%%%%%%%%%
\begin{figure*}
  \centering
  \includegraphics[width=.31\linewidth]{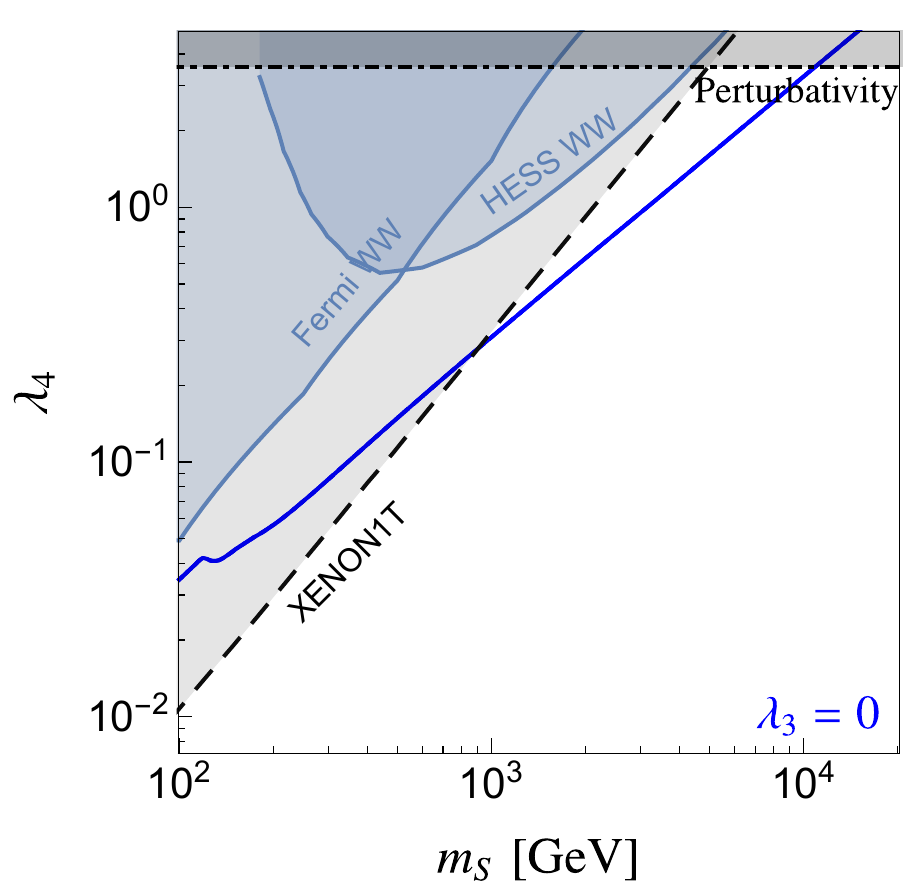} 
  \includegraphics[width=.31\linewidth]{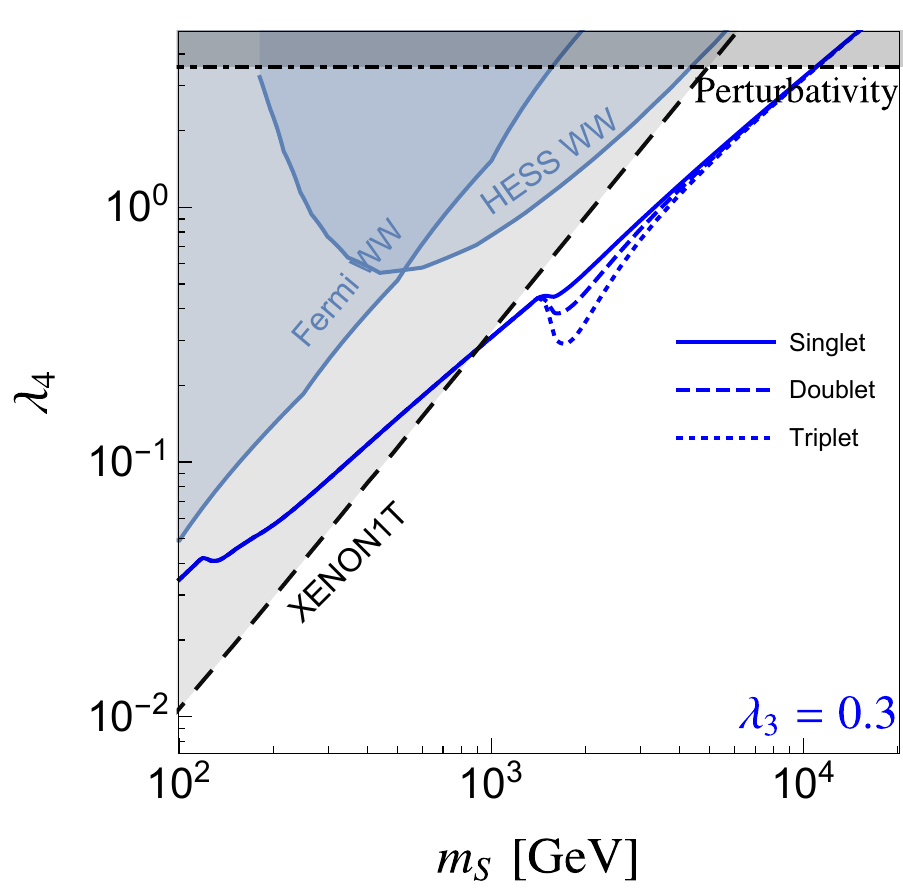} 
  \includegraphics[width=.31\linewidth]{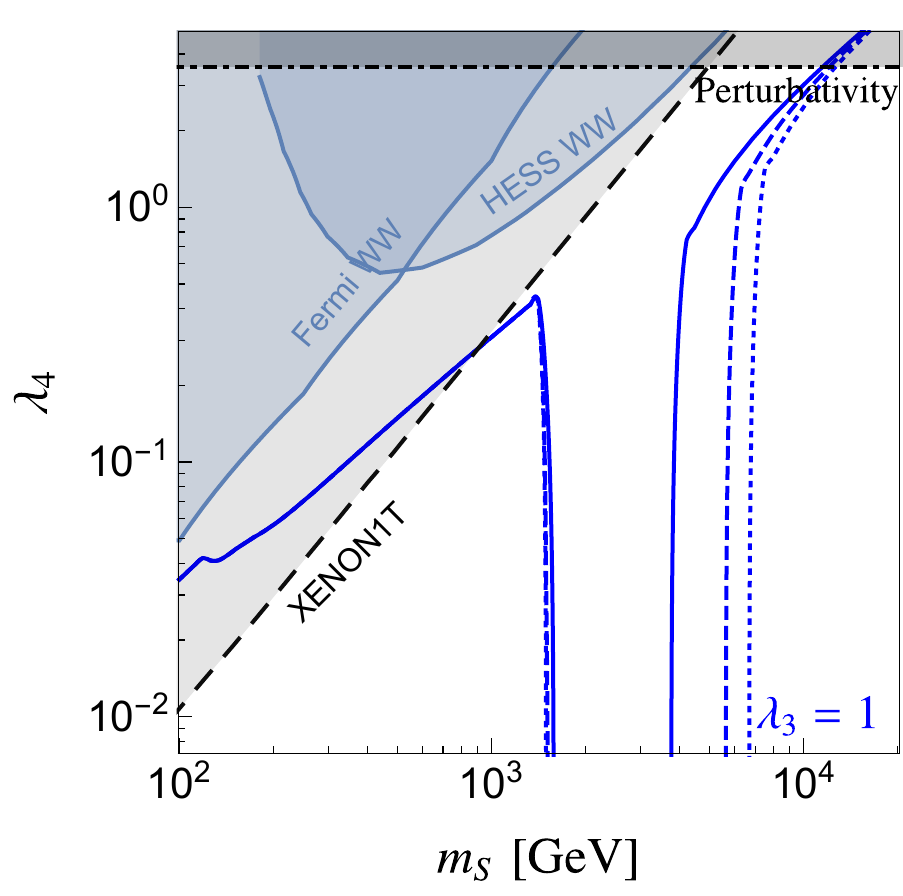}
\caption{\small \sl The values of the Higgs portal coupling $\lambda_4$ needed to explain the DM abundance are plotted in blue as a function of the DM mass $m_S$. Results for singlet ($S_1$, $\widetilde{S}_1$), doublet ($R_2$, $\widetilde{R}_2$) and triplet ($S_3$) LQ states are shown, with $m_\Delta=1.5~\mathrm{TeV}$. These results are largely independent on the hypercharge of the LQ state, depending only on the $SU(2)_L$ representation. Excluded regions due to Xenon1T and perturbativity are depicted by the gray-shaded areas.  The blue-shaded regions are excluded by indirect constraints by HESS~\cite{1607.08142} and FermiLAT~\cite{1503.02641}. Note that in the right-most plot there are forbidden intervals of $m_S$, as there the annihilation cross-section is too large due to the large value of $\lambda_3$.}
\label{fig:lamb3-sigma}
\end{figure*}
%%%%%%%%%%%%%%%%%%%
%%%%%%%%%%%%%%%%%%%

\subsubsection{LHC direct searches} LQs can be produced in pairs at the LHC via gluon fusion and decay into quark-lepton pairs. Searches have been performed at ATLAS and CMS for several possibilities of final states, cf.~e.g.~\cite{Diaz:2017lit,Angelescu:2018tyl,Angelescu:2021lln} for a compilation of the latest LHC bounds. For instance, for LQs decaying into $t\mu$ final states, we find that the most stringent upper limits on LQ masses are $1420~(950)$~GeV for $\mathcal{B}(\Delta\to t\mu)=1~(0.5)$~\cite{Aad:2020jmj}. Similar or weaker limits are obtained for other possible final states. In what follows, we will conservatively take $m_{\Delta} \geq 1.5$~TeV, allowing us to safely avoid existing LHC direct-search limits.

%%%%%%%%
%%%%%%%%
%%%%%%%%
\section{A Viable Scenario with Heavy Dark Matter}
\label{sec:results}
%%%%%%%%
%%%%%%%%
%%%%%%%%

We assess the viability of possible DM scenarios by combining the phenomenological constraints presented in the previous Section. For DM lighter than the weak scale, severely constrained by XENON1T as illustrated in Fig.~\ref{fig:direct-detection}, annihilations to SM fermions via the Higgs boson exchanged in the $s$-channel are dominant. This case is potentially interesting because LQ Yukawas, motivated by discrepancies observed in $B$-meson decays (see e.g.~Ref.~\cite{Buttazzo:2017ixm,Angelescu:2018tyl} and references therein) and in the anomalous magnetic moment of the muon~\cite{Dorsner:2019itg,Cheung:2001ip,Dorsner:2020aaz}, could alter the na\"ive expectation for the annihilation rates. More precisely, these LQ flavour effects would come into play via modification of the Higgs Yukawas contributing to $SS\to h^\ast \to \ell\ell$ via a chiral enhancement of the amplitudes ($\propto m_t/m_\ell$) which lifts up the cross section significantly compared to the standard Higgs portal.

In order to show that the low-mass DM regime is not viable, we take the LQ $R_2=(\mathbf{3},\mathbf{2},7/6)$ as benchmark since the chiral enhancement of the leptonic amplitude is possible for this model. If we set the LQ Yukawa couplings to zero, the quartic $\lambda_4$ controls annihilations to SM fermions in the low DM mass region, as depicted in~Fig.~\ref{fig:T-avg}. This is also true if we switch on the Yukawa couplings, with a chirally enhanced contribution to the amplitude for annihilation to lepton pairs $\ell^+ \ell^-$ proportional to $y_{L}^{t \ell}\, y_{R}^{t \ell}$ (see Eq.~\eqref{eq:ModYuk} of App.~\ref{app:AnnihilationXSec}). We look for the values of $\lambda_4$ necessary to explain the observed relic density for different values of $|y_{L}^{t \ell}\, y_{R}^{t \ell}|$ within the perturbative regime, and we compare them with bounds on $\lambda_4$ stemming from the invisible Higgs width in Eq.~\eqref{eq:hinv-constr} and the XENON1T limits depicted in Fig.~\ref{fig:direct-detection}. In the low DM mass region, there is no value of $\lambda_4$ compatible with relic density and not excluded by the constraints mentioned above. The couplings $y_{t\ell}^L\,y_{t\ell}^R$ provide a non-neglibible decrease in the required value of $\lambda_4$ but their effect cannot bring $\lambda_4$ to the $\mc{O}(10^{-2})$ level which is needed.

In order to be compatible with the severe experimental constraints at low DM mass, we consider values of $m_S$ above the electroweak scale. LQ Yukawas are irrelevant in this DM mass region since they do not impact the main annihilation channels illustrated in Fig.~\ref{fig:T-avg} and \ref{fig:T-avg-2}, and annihilations are controlled by the quartic couplings $\lambda_{3,4,5}$. In what follows, we explore in detail the viability of such a heavy DM scenario. 

\begin{figure*}
  \centering
  \includegraphics[width=.31\linewidth]{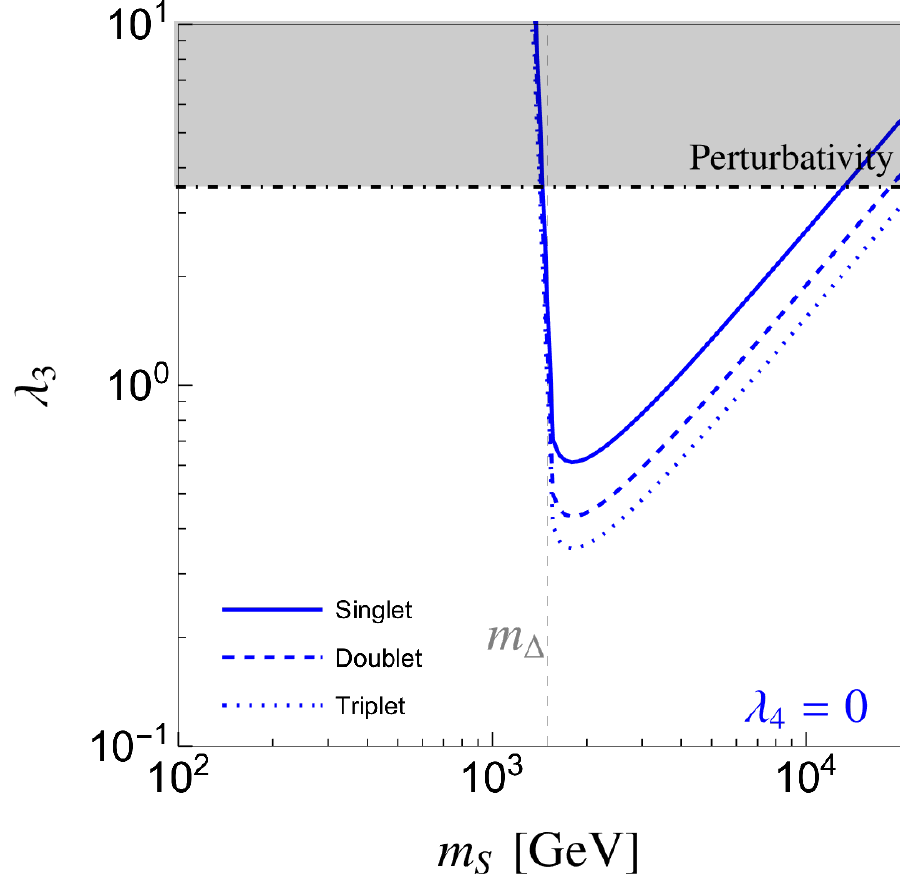}
  \includegraphics[width=.31\linewidth]{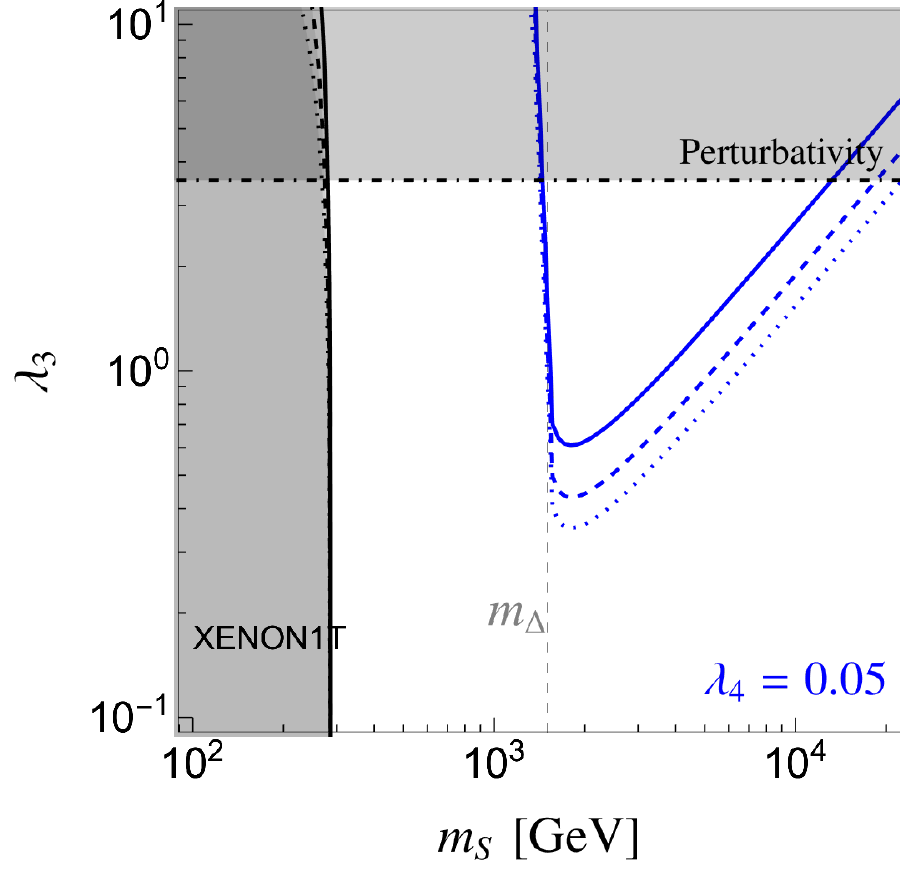}
  \includegraphics[width=.31\linewidth]{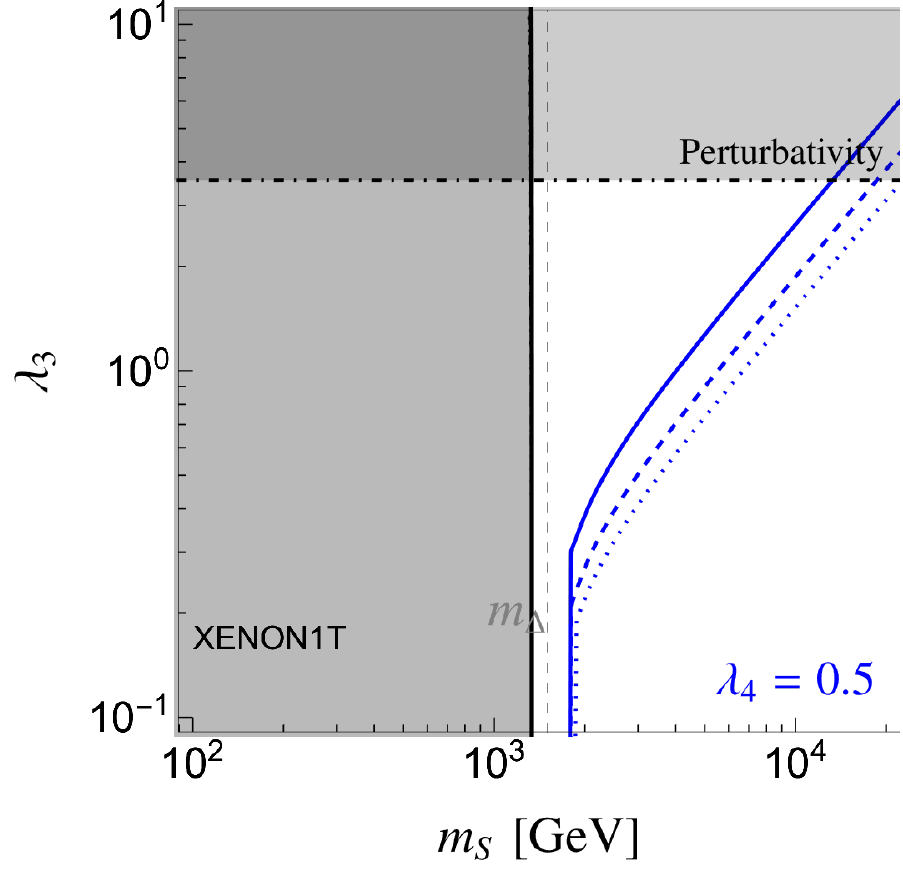}
  \caption{\small \sl The values of the LQ-portal coupling $\lambda_3$ needed to explain the DM abundance are plotted in blue as a function of the DM mass $m_S$. Results for singlet ($S_1$, $\widetilde{S}_1$), doublet ($R_2$, $\widetilde{R}_2$) and triplet ($S_3$) LQ states are shown for $m_\Delta=1.5~\mathrm{TeV}$. DM constraints turn out to be largely independent on the hypercharge of the LQ state, depending only on the $SU(2)_L$ representation. Excluded regions due to Xenon1T and perturbativity are depicted by the gray-shaded areas. Note that below the $\Delta \Delta^*$ threshold the value of $\lambda_3$ becomes very large in the first two plots, whereas in the last plot ($\lambda_4 = 0.5$) there is no solution below a certain mass $m_S$ close to the threshold since the annihilation cross-section is too large.}
  \label{fig:lamb4-relic}
\end{figure*}

We now consider the singlet ($S_1$, $\widetilde{S}_1$), doublet ($R_2$, $\widetilde{R}_2$) and triplet ($S_3$) LQ states $\Delta$ with mass $m_\Delta = 1.5~\mathrm{TeV}$, in agreement with the LHC direct limits from Sec.~\ref{sec:collider-constraints}. The leading impact of the LQ hypercharge ($Y$) is to modify the subleading annihilation channel $SS \to ZZ$, where its contribution is suppressed by $s_W^4$ (cf. Eq.~\eqref{eq:DVVdirectZZ}). For this reason, we observe no noticable effect of $Y$ in Figs.~\ref{fig:lamb3-sigma} and \ref{fig:lamb4-relic}.
Furthermore, we neglect the LQ Yukawa couplings since they have negligible impact on DM phenomenology in this mass region. We show in Fig.~\ref{fig:lamb3-sigma} a slice of the parameter space in the $(m_S, \lambda_4)$ plane by fixing the other couplings to benchmark values, namely $\lambda_5=0$ and $\lambda_3\in \lbrace 0, 0.3,1\rbrace$. Blue lines correspond to the values of $\lambda_4$ and $m_S$ needed to explain the DM relic abundance. The excluded regions stem from perturbativity constraints, Higgs observables, and direct and indirect DM searches (depicted by the shaded regions). For small values of the couplings $\lambda_3$, these scenarios resemble the usual Higgs portal scenario~\cite{Silveira:1985rk,hep-ph/0011335,1009.5377}, with the exception that LQs contribute to the DM annihilation into SM gauge bosons via their virtual effects. In this case, the dominant DM annihilation mechanism are $SS\to VV$, as one could already infer from Fig.~\ref{fig:T-avg}. For larger values of $\lambda_3$, annihilations $SS\to \Delta \Delta^*$ become largely dominant and the relic density depends to first approximation only on $\lambda_3$, explaining the almost vertical lines in the bottom plot of Fig.~\ref{fig:lamb3-sigma}. As we can see from this Figure, it is possible to find a viable scenario for all benchmark couplings if and only if $m_S\gtrsim m_\Delta$.  

%%%%%%%%%%%%%%%%%%%
%%%%%%%%%%%%%%%%%%%
%%%%%%%%%%%%%%%%%%%
%%%%%%%%%%%%%%%%%%%

%%%%%%%%%%%%%%%
%%%%%%%%%%%%%%%%
\begin{figure*}[!t]
  \centering
  \includegraphics[width=.9\linewidth]{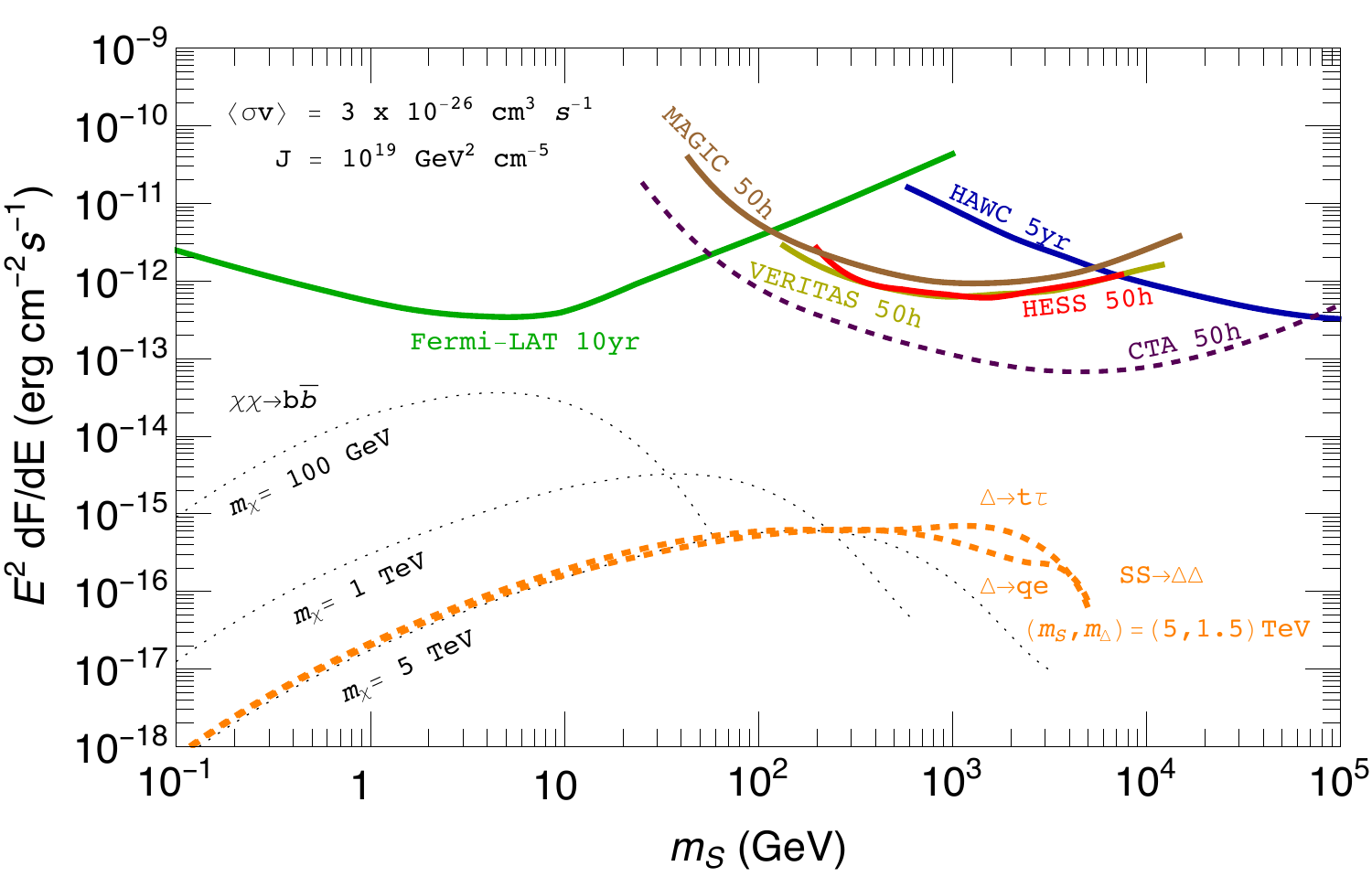}
  \caption{\small \sl Flux of gamma rays from DM annihilations to LQs and subsequent LQ decays to SM particles (dashed orange lines). We consider both LQ decays to third generation fermions ($t \tau$) as well as lighter fermions ($q e$). The annihilation cross section reproduces the relic abundance via thermal freeze-out, and we consider a dwarf galaxy with the J-factor provided in the legend. We report for comparison also the expected spectrum for a WIMP candidate annihilating to $b \bar{b}$ pair for different masses (dotted black lines). The sensitivity lines for current (solid) and future (dashed) experiments are taken from reference~\cite{Rico:2020vlg}.}
  \label{fig:indirectdetection}
\end{figure*}
%%%%%%%%%%%%%%%
%%%%%%%%%%%%%%%

To further explore the regime where $SS\to \Delta \Delta^*$ acts as the main annihilation channel, we plot in Fig.~\ref{fig:lamb4-relic} the allowed DM parameters in the  $(m_S, \lambda_3)$ plane by keeping $\lambda_5=0$ and fixing this time $\lambda_4\in \lbrace 0, 0.05, 0.5\rbrace$. For the pure LQ portal ($\lambda_4 = 0$) shown in the left panel, we see that a coupling $\lambda_3 \gtrsim 0.5$ is needed to reproduce the DM relic abundance. For larger values of $\lambda_4$, both $SS\to \Delta \Delta^*$ and $SS\to VV$ annihilation channels become relevant, and smaller values of $\lambda_3$ are needed as a consequence. Direct detection constraints are mostly sensitive to $\lambda_4$ and practically insensitive to $\lambda_3$. There are two potential constraints on the LQ portal coupling $\lambda_3$: indirect detection and perturbativity. The former is sub-dominant, as we quantify in the next paragraph, and the only meaningful constraint on $\lambda_3$ arises from the perturbative bounds.

%%%%%%%%%%%%%%%%
%%%%%%%%%%%%%%%%
\begin{figure*}[!t]
	\centering
        \begin{tabular}{lr}
          \includegraphics[width=.49\linewidth]{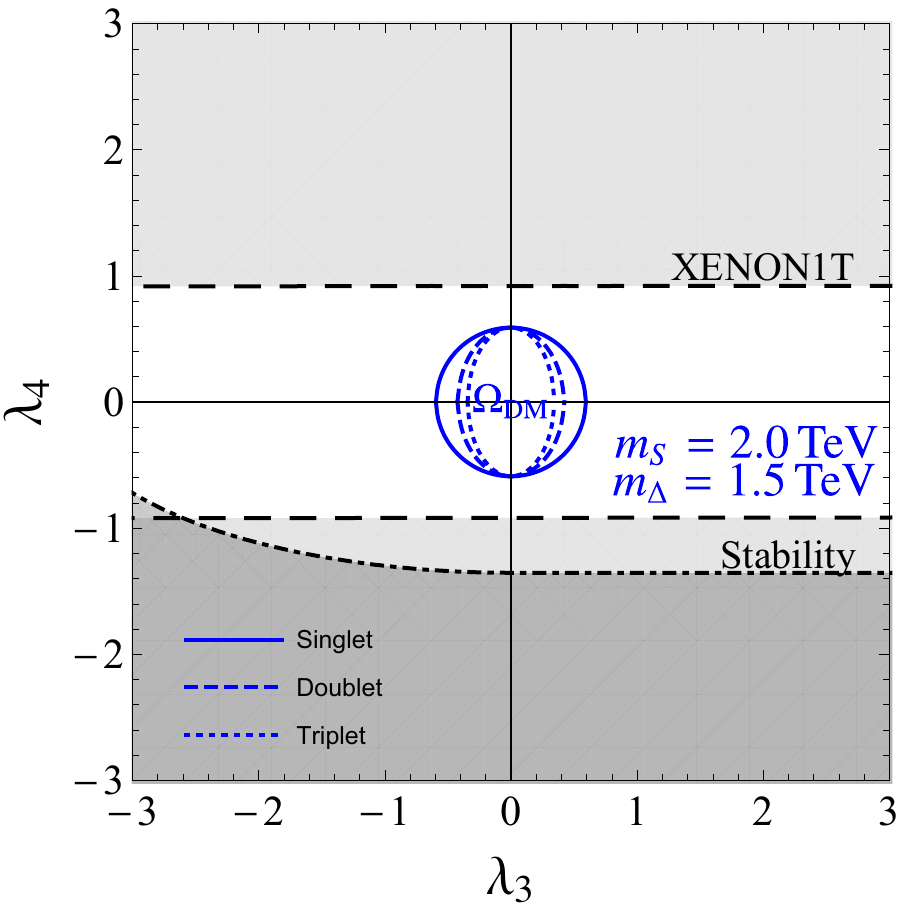} & $\;\;$ \includegraphics[width=.49\linewidth]{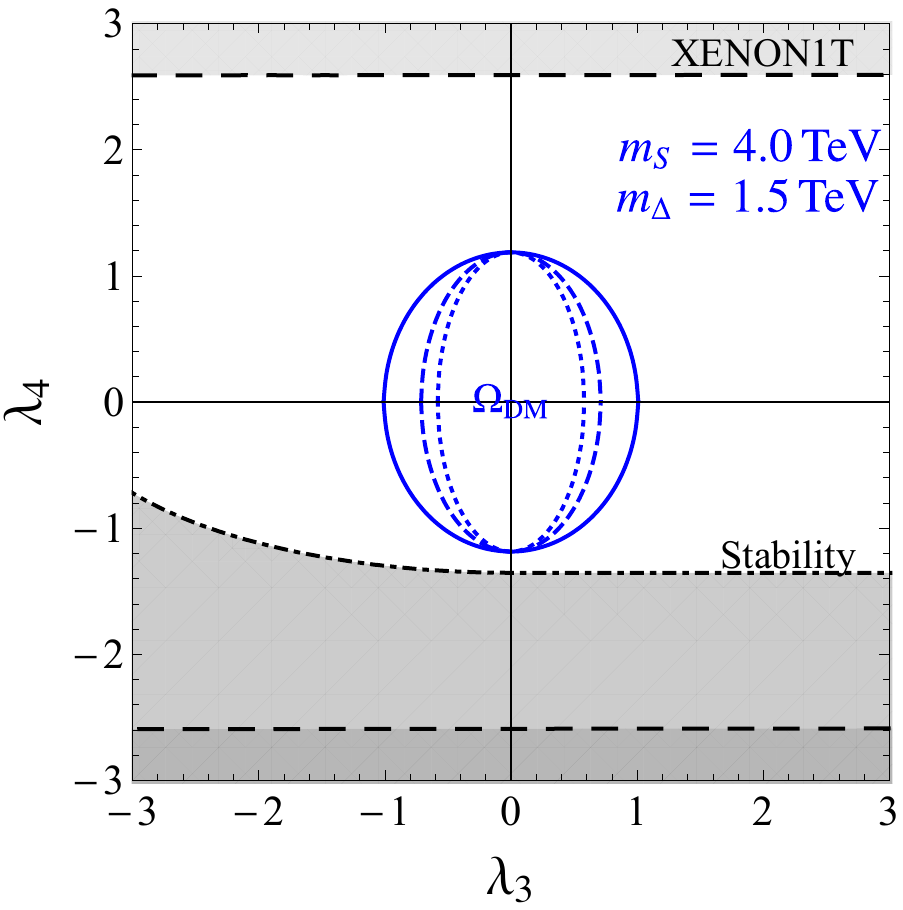}
        \end{tabular}		
	\caption{\small \sl Constraints on the LQ-portal $\lambda_3$ and the Higgs-portal $\lambda_4$ couplings imposed by the relic abundance (reproduced exactly along the blue line), direct detection due to Xenon 1T (exclusions with a dashed border) and the potential stability constraints (exclusion with a dash-dotted border). Results for singlet ($S_1$, $\widetilde{S}_1$), doublet ($R_2$, $\widetilde{R}_2$) and triplet ($S_3$) LQ states are shown, with two representative values of $m_S$ above $m_{\Delta}=1.5$~TeV. DM constraints turn out to be largely independent on the hypercharge of the LQ state, depending only on the $SU(2)_L$ representation.
	}
	\label{fig:l3l4}
\end{figure*}
%%%%%%%%%%%%%%%
%%%%%%%%%%%%%%%%

We compute the gamma-ray spectrum produced by the cascade annihilations, first with DM annihilating to LQ final state $SS\to \Delta \Delta^*$ and then subsequent LQ decays, and we show our results in Fig.~\ref{fig:indirectdetection}. We fix the annihilation cross section to the thermal relic value, and we consider a dwarf galaxy with a J-factor equal to $J = 10^{19} \, {\rm GeV}^2 \, {\rm cm}^{-5}$. The dashed orange lines provide the photon flux for the representative spectrum $(m_S, m_\Delta) = (5, 1.5) \, {\rm TeV}$ and for two possible LQ decays: third generation fermions ($\Delta \rightarrow t \tau$) and lighter fermions ($\Delta \rightarrow q e$). The larger top mass is responsible for a slight enhancement in the UV tail of the spectrum. We report for comparison also WIMP spectra for annihilations to $b \bar{b}$ final state for different DM masses. The only region where the LQ portal spectrum differs slightly from the one induced by annihilations of a WIMP with the same mass is in the UV tail. We show in the same figure the sensitivity curves for different instruments, as collected in Ref.~\cite{Rico:2020vlg}. The Fermi bounds for DM mass around the TeV scale in the LQ portal benchmark are for all purposes the same as the one for a WIMP with the same mass; thermal relics are not excluded. For instruments reaching maximum sensitivity around the TeV scale, the slight enhancement in the UV tail could slightly affect the bounds for WIMPs but they should not spoil the picture completely; these instruments are still quite far from the thermal relic line and therefore thermal relics are also not excluded by them. We conclude that current indirect detection constraints are never the most stringent ones for the LQ portal case. We also compare the sensitivity of the future CTA telescope with the signal predicted in our model.

We visualize the interplay among the different constraints in Fig.~\ref{fig:l3l4} where we fix $m_S$ to two benchmark values in the TeV range, and we vary the LQ and Higgs portal-couplings $\lambda_3$ and $\lambda_4$. We notice how an improvement of the direct-dection constraints by a factor of $\approx 4$ is needed to start probing $\lambda_3$ in this range of DM masses. Moreover, stability constraints (Eq.~\eqref{eq:tetrahedronCop}) turn out to be meaningful for negative values of $\lambda_4$.

Finally, we explore the extent to which our conclusions remain valid when the LQ mass is increased above our benchmark value $m_\Delta=1.5$~TeV. To this purpose, we focus on the pure LQ portal, with $\lambda_4$ set to zero. The values of $\lambda_3$ needed to reproduce the relic density are then shown in Fig.~\ref{fig:pert} in the plane $m_\Delta$~vs.~$m_S$ for singlet, doublet and triplet LQ states. The boundary of the allowed region is determined by the perturbativity constraints, which imply the following upper limit on both the LQ and DM masses
%%%%%%%%%%%%%%%%%%
\begin{equation}
\label{eq:pert-bound}
 m_\Delta < m_S < \mathcal{O}(10~\mathrm{TeV})\,.
\end{equation}
%%%%%%%%%%%%%%%%%%%

\noindent The perturbativity contours for the doublet and triplet LQ states are broader than the corresponding singlet contour. This is a consequence of higher LQ multiplicity, which increases the annihilation cross section and slightly relaxes the upper limit Eq.~\eqref{eq:pert-bound}. However, the main phenomenoligcal features remain very similar for all scenarios.

\section{Implications for Flavour Anomalies}
\label{sec:FA}

In this Section, we discuss the implications of the viable DM scenarios outlined in Sec.~\ref{sec:results} to the discrepancies observed in $B$-meson decays and the muon $g-2$. Both types of discrepancies suggest that LQs could exist at the TeV scale with $\mathcal{O}(1)$ Yukawa couplings to the SM fermions. Therefore, it is a natural question if these anomalies could be accommodated in a scenario that would also explain the DM abundance via the mechanism discussed in Sec.~\ref{sec:results}.

%%%%%%%%%%%%%%%%%%%
%%%%%%%%%%%%%%%%%%%
\begin{figure}[!t]
  \centering
  \includegraphics[width=.95\linewidth]{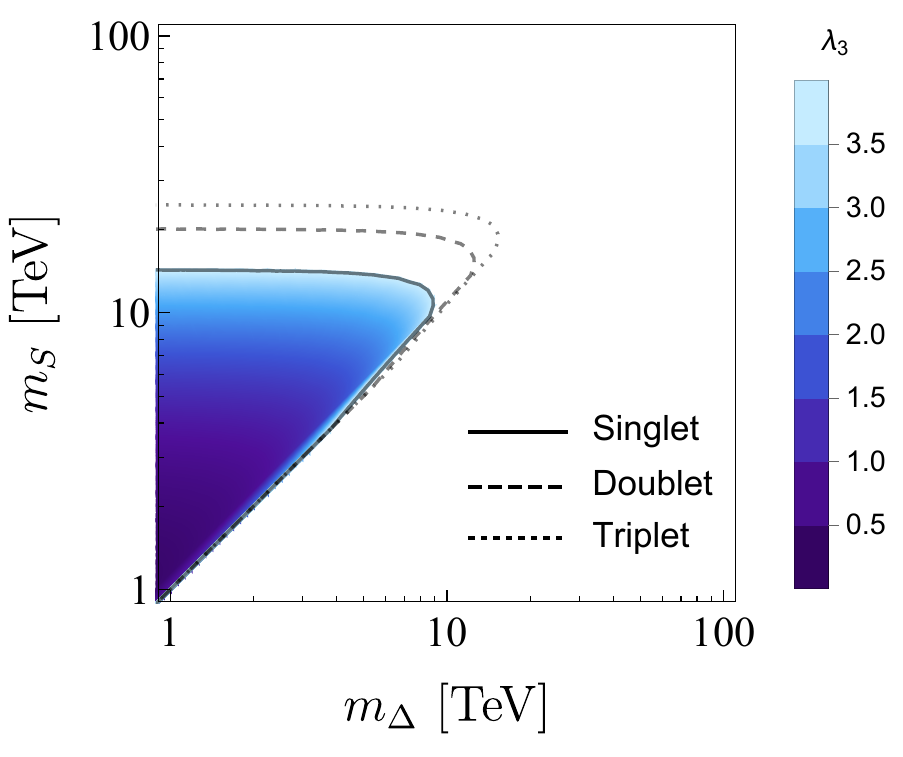}
  \caption{\small \sl Dependence of the required $\lambda_3$ coupling on the LQ mass $m_\Delta$ and the DM mass $m_S$ in the pure LQ-portal scenario with $\lambda_4=0$, imposed by the relic abundance. The contours show the values of the LQ-portal $\lambda_3$ needed to reproduce the relic abundance exactly in the case of a singlet LQ state, whereas the contours are cut-off at the perturbativity bound $\lambda_3 = \sqrt{4\pi}$. Dashed and dotted contours denote the perturbative regions for the cases of doublet and triplet LQ states, respectively.
}
  \label{fig:pert}
\end{figure}
%%%%%%%%%%%%%%%%%%%
%%%%%%%%%%%%%%%%%%%

To answer the question raised above, we will remind the reader which scalar LQ states can successfully accommodate each of these anomalies and we will derive the ranges of masses $m_\Delta$ that are compatible with them~\cite{DiLuzio:2017chi,Allwicher:2021jkr}. These values will be compared to the upper limits on $m_\Delta$ obtained in Fig.~\ref{fig:pert} in order to provide a viable DM scenario. As already anticipated in Sec.~\ref{sec:results}, the specific values of the LQ Yukawas that are fixed at low-energies have little impact on the DM abundance, in such a way that the only common parameter for flavor and DM phenomenology is indeed the LQ mass, which is explored in the following.

\

\paragraph*{$\bullet$~{$R_{K^{(\ast)}}$}\,:} Several discrepancies in exclusive $B$-meson decays based on the transition $b\to s\ell\ell$ have been recently observed by LHCb. Most importantly, the Lepton Flavour Universality (LFU) ratio~\cite{Aaij:2021vac}
%%%%%%%%%%%%%%%%
\begin{align}
\begin{split}
 R_{K}^{[1.1,6]}\Big{\vert}_\mathrm{exp} &= \dfrac{\mathcal{B}(B\to K\mu \mu)}{\mathcal{B}(B\to K ee)}\\[0.3em]
 &=0.846^{+0.042}_{-0.039}\,^{+0.013}_{-0.012}\,,
\end{split}
\end{align}
%%%%%%%%%%%%%%%
integrated in the di-lepton invariant-mass bin $q^2\in (1.1, 6)~\mathrm{GeV}^2$, turns out to be $3.1\,\sigma$ below the precise SM prediction, $R_K^\mathrm{SM}=1.00(1)$~\cite{Hiller:2003js,Isidori:2020acz}. Deviations from the SM predictions have also been observed in similar LFU tests with $B\to K^\ast \ell\ell$ decays~\cite{Aaij:2017vbb}. The combination of these results with the current experimental average of the $B_s\to \mu\mu$ branching fraction~\cite{CMS:2020rox,LHCbNEW}
%%%%%%%%%%%%%%%
\begin{equation}
\overline{\mathcal{B}}(B_s\to\mu\mu)^\mathrm{exp}=2.85(33)\times 10^{-9}\,,
\end{equation}
which is also slightly below the clean SM prediction $\overline{\mathcal{B}}(B_s\to\mu\mu)^\mathrm{SM}=3.66(14)\times 10^{-9}$~\cite{Beneke:2019slt}, amounts to a combined deviation of $4.6\,\sigma$ from the SM predictions~\cite{Angelescu:2021lln} (see also~Ref.~\cite{Geng:2021nhg,Altmannshofer:2021qrr,Cornella:2021sby}). 

The preferred scenario to explain these deviations requires a purely left-handed operator,
%%%%%%%%%%%%%%%
\begin{equation}
\label{eq:leff-bsmumu}
 \mathcal{L}_\mathrm{eff} \supset \dfrac{C_L^{\mu\mu}}{v^2} \big{(}\bar{s}_L \gamma^\mu {b}_L\big{)}\big{(}\bar{\mu}_L \gamma_\mu \mu_L\big{)}+\mathrm{h.c.}\,,
\end{equation}
%%%%%%%%%%%%%%%
with the effective coefficient $C_L^{\mu\mu}$ in the following range~\cite{Angelescu:2021lln},
%%%%%%%%%%%%%%%
\begin{equation}
\label{eq:fit-bsmumu}
C_L^{\mu\mu} = (4.1\pm 0.9)\times 10^{-5}\,.
\end{equation}
%%%%%%%%%%%%%%%

\noindent Among the scalar LQs listed in Sec.~\ref{sec:lq-classification}, only the scalar triplet $S_3=(\overline{\mathbf{3}},\mathbf{3},1/3)$ can induce a nonzero value of $C_L^{\mu\mu}$ at tree-level~\cite{Hiller:2014yaa,Angelescu:2021lln},
%%%%%%%%%%%%%%%
\begin{equation}
\label{eq:bsmumu-c9c10}
C_L^{\mu\mu} = \dfrac{v^2}{m_{S_3}^2}\big{(}y^L_{S_3}\big{)}_{b\mu} \big{(}y^L_{S_3}\big{)}_{s\mu}^\ast\,,
\end{equation}
%%%%%%%%%%%%%%%

\noindent where we remind the reader that the Yukawas $y^L_{S_3}$ are defined in Eq.~\eqref{eq:yuk-S3}. By combining Eq.~\eqref{eq:fit-bsmumu} and \eqref{eq:bsmumu-c9c10}, it is straightforward to conclude that
%%%%%%%%%%%%%%%
\begin{equation}
\dfrac{\big{(}y^L_{S_3}\big{)}_{b\mu} \big{(}y^L_{S_3}\big{)}_{s\mu}^\ast}{m_{S_3}^2} \simeq \dfrac{1}{(38~\mathrm{TeV})^2}\,.
\end{equation}
%%%%%%%%%%%%%%%

\noindent This constraint implies an upper bound on $m_{S_3}$ which is compatible with, but much less strict than the bound obtained in Fig.~\ref{fig:pert} from the requirement of reproducing the correct DM relic density via the mechanism discussed in this paper. The bound on $m_{S_3}$ from relic density becomes significantly stronger for small mass $m_S$.

\

\paragraph*{$\bullet$~{$R_{D^{(\ast)}}$}\,:} Several discrepancies from the SM predictions have also been observed in LFU tests for the transition $b\to c\ell \nu$,
%%%%%%%%%%%%%%%%%%%
\begin{align}
  R_{D^{(\ast)}} &= \dfrac{\mathcal{B}(B\to D^{(\ast)} \tau \bar{\nu})}{\mathcal{B}(B\to D^{(\ast)}  l \bar{\nu})}\,,\quad (l=e,\mu)
\end{align}
%%%%%%%%%%%%%%%%%%%
The current experimental averages of LHCb~\cite{Aaij:2015yra,Aaij:2017deq} and the $B$-factories \cite{Lees:2013uzd,Huschle:2015rga,Hirose:2016wfn,Hirose:2017dxl,Belle:2019rba} measurements are~\cite{Amhis:2019ckw}
%%%%%%%%%%%%%%%%%%%
\begin{align}
\begin{split}
  R_{D}^\mathrm{exp} &= 0.340\pm 0.030\,,\\[0.35em]
  R_{D^\ast}^\mathrm{exp} &=0.295\pm 0.013\,,
\end{split}
\end{align}
%%%%%%%%%%%%%%%%%%%
which turns out to be $\approx 1.3\,\sigma$ and $\approx 3.5\sigma$ above the SM predictions,
%%%%%%%%%%%%%%%%%%%
\begin{align}
  \begin{split}
  \label{eq:lattice-ff}
R_{D}^\mathrm{SM} &= 0.299\pm 0.003\,,\\[0.35em] 
  R_{D^\ast}^\mathrm{SM} &=0.2484\pm 0.0013\,,
  \end{split}
\end{align}
%%%%%%%%%%%%%%%%%%%
\noindent which are obtained by combining the latest lattice QCD results for the $B\to D^{(\ast)}$~\cite{Aoki:2019cca,Bazavov:2021bax} form factors at nonzero recoil with the $B\to D^{(\ast)}l \bar{\nu}$ ($l=e,\mu$) differential decay rates measured experimentally, see~\cite{Gambino:2020jvv,Bazavov:2021bax} and references therein.

The observed deviations in $R_{D^{(\ast)}}$ can also be interpreted by means of an effective field theory,
%%%%%%%%%%%%%%%%%%%
\begin{align}
 \mathcal{L}_\mathrm{eff} &\supset -2\sqrt{2}G_F V_{cb} \sum_a g_a \,\mathcal{O}_a+\mathrm{h.c.}\,,
\end{align}
%%%%%%%%%%%%%%%%%%
where $g_a$ are effective couplings, defined at the renormalization scale $\mu=m_b$, and the relevant effective operators are
%%%%%%%%%%%%%%%%%%%
\begin{align}
%\begin{split}
 \mathcal{O}_{V_L} &= (\bar{c}_{L}\gamma_\mu {b}_{L}) (\bar{\ell}_L\gamma^\mu\nu_L)\,,\\[0.35em]
%  \mathcal{O}_{V_R} &= (\bar{c}_{R}\gamma_\mu {b}_{R}) (\bar{\ell}_L\gamma^\mu\nu_L)\,,\\[0.35em]
 \mathcal{O}_{S_L} &= (\bar{c}_{R} b_{L})(\bar{\ell}_R \nu_L)\,,\\[0.35em]
% \mathcal{O}_{S_R} &= (\bar{c}_{L} b_{R})(\bar{\ell}_R \nu_L)\,,\\[0.35em]
 \mathcal{O}_{T} &= (\bar{c}_R \sigma_{\mu\nu}b_L)(\bar{\ell}_R \sigma^{\mu\nu} \nu_L)\,,
% \end{split}
\end{align}
%%%%%%%%%%%%%%%%%%
as well as $\mathcal{O}_{V_R}$ and $\mathcal{O}_{S_R}$ which are obtained from the ones above by flipping the chirality of the quark fields. Differently from the previous case (see Eq.~\eqref{eq:leff-bsmumu}), there is more than one viable effective scenario induced by LQs that can explain the anomalies in $R_{D^{(\ast)}}$. The simplest viable scenario is to consider once again a purely left-handed operator,
%%%%%%%%%%%%%%%%%%%
\begin{equation}
 g_{V_L} = 0.084(20)\,,
\end{equation}
%%%%%%%%%%%%%%%%%%
\noindent where we have updated results from Ref.~\cite{Angelescu:2021lln} by considering the latest lattice QCD results for the $B\to D^\ast$ form factors~\cite{Bazavov:2021bax}. Another option is to consider the effective scenarios $g_S = \pm 4\, g_T$, at the matching scale $\mu = m_\Delta\approx 1~\mathrm{TeV}$, which can also perfectly describe current data~\cite{Angelescu:2021lln}. These relations become $g_{S_L}(m_b) \approx -8.5 g_T(m_b)$ and $g_{S_L}(m_b) \approx +8.1 g_T(m_b)$ at $\mu=m_b$, respectively, after accounting for the RGE effects~\cite{Gonzalez-Alonso:2017iyc}. These scenarios can provide a viable explanation for the $b\to c \tau \bar{\nu}$ anomalies for real and purely imaginary couplings, respectively~\cite{Angelescu:2021lln} (see also Ref.~\cite{Sakaki:2013bfa,Becirevic:2018afm}).
%%%%%%%%%%%%%%%%%%%
%\begin{align}
% g_{S_L}(m_b) &= -8.5\,g_T(m_b) =0.16\pm 0.04\,,\\[0.35em]
% g_{S_L}(m_b) &= +8.1\,g_T(m_b) =i\,(0.49\pm 0.07) \,.
%\end{align}
%%%%%%%%%%%%%%%%%%

There are only two scalar LQs that can predict the allowed values of effective couplings at low-energies, while being consistent with various flavour and LHC constraints~\cite{Angelescu:2021lln}. The first scenario that can be matched to the viable effective scenarios described above is $S_1=(\overline{\mathbf{3}},\mathbf{1},1/3)$, which can predict nonzero values for both $g_{V_L}$ and $g_{S_L}=- 4 g_T$ via the products of couplings $y^L_{S_1}\, y^{L\,\ast}_{S_1}$ and $y^L_{S_1} \, y^{R\,\ast}_{S_1}$, respectively. The needed couplings to explain the anomalies can then be either
%%%%%%%%%%%%%%%%
\begin{align}
\label{eq:S1-LH}
\dfrac{\big{(}y^L_{S_1}\big{)}_{b\tau}\big{(}V\,y^L_{S_1}\big{)}_{c\tau}^\ast}{m_{S_1}^2} &\simeq +\dfrac{1}{(2.1~\mathrm{TeV})^2}\,,
\end{align}
%%%%%%%%%%%%%%%%
or
%%%%%%%%%%%%%%%%
\begin{align}
\label{eq:S1-RH}
\dfrac{\big{(}y^L_{S_1}\big{)}_{b\tau}\big{(}y^R_{S_1}\big{)}_{c\tau}^\ast}{m_{S_1}^2} &\simeq -\dfrac{1}{(2~\mathrm{TeV})^2}\,,
\end{align}
%%%%%%%%%%%%%%%%
which both require LQ massses in the $\mathcal{O}(\mathrm{TeV})$ range.

The second viable scenario is $R_2=(\mathbf{3},\mathbf{2},7/6)$, which can explain current data provided there is an imaginary phase in the LQ Yukawas~\cite{Sakaki:2013bfa,Angelescu:2021lln}, 
%%%%%%%%%%%%%%%%
\begin{align}
\label{eq:yuk-R2RD}
\dfrac{\big{(}y^L_{R_2}\big{)}_{c\tau}\big{(}y^R_{R_2}\big{)}_{b\tau}^\ast}{m_{R_2}^2} &\simeq \pm \dfrac{i}{(1.1~\mathrm{TeV})^2}\,.
\end{align}
%%%%%%%%%%%%%%%%

\noindent which is also in the $\mathcal{O}(\mathrm{TeV})$ range.

Therefore, by inspecting Eq.~\eqref{eq:S1-LH}--\eqref{eq:S1-RH} and \eqref{eq:yuk-R2RD}, we find that the range of LQ masses needed to explain $R_{D^{(\ast)}}$ turns out to be very similar to the one needed to explain the DM relic abundance, see Fig.~\ref{fig:pert}. Note, also, that these masses and LQ couplings are fully consistent with high-$p_T$ constraints and with other flavor constraints, as recently analyzed in Ref.~\cite{Angelescu:2021lln}.

\

\paragraph*{$\bullet$~{$(g-2)_\mu$}\,:} Lastly, we discuss the impact of scalar LQs to the $4.2\,\sigma$ discrepancy observed between the experimental determinations of $a_\mu=(g-2)_\mu/2$~\cite{Bennett:2006fi,Abi:2021gix} and the SM prediction from the Muon $g-2$ theory initiative~\cite{Aoyama:2020ynm}, 
%%%%%%%%%%%%%%%%
\begin{align}
 \Delta a_\mu = a_\mu^\mathrm{exp}-a_\mu^\mathrm{SM} = (251\pm 59)\times 10^{-11}\,.
\end{align}
%%%%%%%%%%%%%%%

\noindent This discrepancy is comparable in size to the SM electroweak contributions. Therefore, it is only possible to explain it through LQs contributions with $\mathcal{O}(\mathrm{TeV})$ masses if a chirality-enhancement mechanism takes place~\cite{Feruglio:2018fxo,Aebischer:2021uvt,Fajfer:2021cxa,Crivellin:2021rbq}. Such an enhancement can be induced by LQs that couple simultaneously $\mu_L$ and $\mu_R$ to a heavy fermion, which is typically the top quark, producing an enhancement $\propto m_t/m_\mu$~\cite{Cheung:2001ip}.~\footnote{Charm-quark loops are also sizably enhanced~\cite{Kowalska:2018ulj}, but these contributions are tightly constrained by $pp\to\mu\mu$  at high-$p_T$~\cite{Angelescu:2021lln}. Bottom-quark loops are only possible via the mixing of two scalar LQs~\cite{Dorsner:2019itg}.} There are only two LQ states that satisfy this criteria, namely $S_1=(\overline{\mathbf{3}},\mathbf{1},1/3)$ and $R_2=(\mathbf{3},\mathbf{2},7/6)$. The needed couplings for $S_1$ read
%%%%%%%%%%%%%%%%%%
\begin{equation}
\label{eq:lq-gminus2-S1}
\dfrac{\big{(} y^L_{S_1}\big{)}_{t\mu} \big{(}y^{R}_{S_1} \big{)}_{t\mu}^\ast}{m_{S_1}^2} \simeq +\dfrac{1}{(32~\mathrm{TeV})^2}\,,
\end{equation}
%%%%%%%%%%%%%%%%%%%
whereas for $R_2$,
%%%%%%%%%%%%%%%%%%
\begin{equation}
\label{eq:lq-gminus2-R2}
\dfrac{\big{(}y^L_{R_2}\big{)}_{t\mu}\big{(}y^{R}_{R_2}\big{)}_{t\mu}^\ast}{m_{R_2}^2} \simeq -\dfrac{1}{(37~\mathrm{TeV})^2}\,,
\end{equation}
%%%%%%%%%%%%%%%%%%%
where, for simplicitly, we have kept only the chirality-enhanced contributions and set the LQ masses to $10~\mathrm{TeV}$ in the logarithms. In this case, we find that reproducing the observed DM relic density imposes a much stricter constain than the muon $g-2$ which is sensitive to very large LQ masses. Note, also, that the constraints obtained in Eq.~\eqref{eq:lq-gminus2-S1} and \eqref{eq:lq-gminus2-R2} are much stricter than the ones derived from other loop observables such as $Z\to\mu\mu$~\cite{Arnan:2019olv,Crivellin:2020mjs}. Finally, it is worth stressing that even though $S_1$ and $R_2$ are also the scalar LQs needed to explain $R_{D^{(\ast)}}$, the simultaneous explanation of both anomalies is tightly constrained by $\tau \to \mu \gamma$ which would also be chirality-enhanced for the needed pattern of Yukawas~\cite{Gherardi:2020qhc}.

\

To summarize this discussion, DM phenomenology and flavour physics are complementary probes of the LQ parameters. DM constraints are practically insensitive to the LQ Yukawas, but they can be used to fix the scalar potential parameters $m_S$, $m_\Delta$ and $\lambda_{3,4,5}$, whereas flavour-physics constraints, at tree-level, are only sensitive to the combinations of Yukawas $|y_{ij}|/m_\Delta$, where $y_{ij}$ denotes a generic LQ Yukawa with flavour indices $i,j$. For this reason, the only connection between flavour and DM arises from the LQ masses, which are bounded in both cases by perturbativity constraints, as shown e.g.~in Eq.~\eqref{eq:pert-bound} for the quartic couplings and discussed above for each of these anomalies. By comparing these upper bounds with Eq.~\eqref{eq:pert-bound}, we see that the limit on $m_\Delta$ derived from DM relic abundance is more constraining than the ones derived from most flavour anomalies, with the only exception of $R_{D^{(\ast)}}$, which is a tree-level process in the SM. Therefore, if any of these anomalies is confirmed in the future, one could easily check from Eq.~\eqref{eq:pert-bound} if the LQ scenario favored from flavour physics could be extended to accommodate DM via the mechanism discussed in this paper.

%%%%%%%%
%%%%%%%%
%%%%%%%%
\section{Conclusion}
\label{sec:conclusion}
%%%%%%%%
%%%%%%%%
%%%%%%%%

The origin of DM is a long standing problem in physics of fundamental interactions. The recently observed discrepancies in the lepton flavour universality tests in $B$ meson decays suggest the existence of LQs. In this paper, we have studied a possible connection between these two open problems. We took a real scalar singlet field $S$ as a DM candidate, and introduced a scalar LQ $\Delta$ with mass above $1\e{TeV}$ and $\mathcal{O}(1)$ Yukawa couplings to the SM fermions which can resolve the flavour anomalies. The only renormalizable interactions of the DM field $S$ with the visible fields are the Higgs and LQ portal-couplings present in the scalar potential.

We have studied the role of the LQ in DM phenomenology, focussing in particular on the portal coupling $\lambda_3 \,|\Delta|^2 S^2$, mass $m_\Delta$, and Yukawa couplings. The latter two parameters also play a central role in flavour anomalies. First, we have introduced the most general scalar potential for the scalar fields of our framework, and we have carefully studied the scalar potential stability conditions which yielded correlated upper bounds on the portal couplings. For the LQ Yukawa sector we stated explicitly all possible representations along with their couplings to the SM fermions. Assuming one scalar LQ is present, we have calculated the DM annihilation rates that depend on the scalar potential parameters and Yukawa couplings. Expressions for cross sections for general LQ representation are provided in App.~\ref{app:AnnihilationXSec}. We feed the Boltzmann equation for the $S$ number density with these cross sections and compute the relic density. Up-to date constraints from direct and indirect searches for DM are also derived and they place important upper bounds on the scalar potential couplings. Collider constraints on $S$ have been taken into account, including the very stringent upper bound on Higgs boson invisible width.

Two scalar potential couplings drive the DM phenomenology -- the Higgs portal $\lambda_4 |H|^2 S^2$ and the LQ portal $\lambda_3 |\Delta|^2 S^2$. It was found and shown in Fig.~\ref{fig:lamb3-sigma} that below the threshold for annihilation into LQs, i.e. $m_S < m_\Delta$, the loop effects of $\lambda_3$ are loop-suppressed and cannot compete with the Higgs portal coupling $\lambda_4$. 
On the other hand, we could observe the impact of large LQ Yukawas boosting $SS \to \ell \ell$ annihilation cross sections ($\ell = \mu,\tau$). This effect is potentially large at low DM mass where $\ell \ell$ channel drives the total annihilation cross section, but the Higgs invisible width is too strong a constraint in this mass region and the relic density comes out too small, despite Yukawa enhancement. At higher $m_S$ the Yukawa couplings do not affect the $S$ annihilation cross section significantly. 

When $m_S > m_\Delta$ the $SS \to \Delta \Delta^*$ channel is open and the phenomenology depends on both Higgs and LQ portals. It turns out that large LQ portal $\lambda_3$ implies small Higgs portal $\lambda_4$ and vice versa. Although both scenarios are equally viable the Higgs portal regime is more prone to direct detection constraints as well as to the stability constraints as shown in Fig.~\ref{fig:l3l4}. Further analysis of the pure LQ portal regime with $\lambda_4=0$ reveals a wide parameter space of $\lambda_3$ and $m_S > m_\Delta \gtrsim 1.5\e{TeV}$. Most importantly, the requirements of stability of the scalar potential and perturbativity set limits both on $m_S$ and $m_\Delta$ to lie below $\mc{O}(10\,\mrm{TeV})$, as shown in Fig.~\ref{fig:pert}.

In summary, flavour and DM aspects of the scalar singlet and scalar LQ model are to a large degree decoupled when $m_S$ is below the threshold for annihilation into LQs, whereas above the threshold we may enter the LQ portal regime where both $S$ and $\Delta$ masses are bounded from above due to the pertubativity constraint. The obtained mass bounds on the LQ are stricter than what would be inferred from most flavour anomalies.

\

\paragraph*{Acknowledgments.} We thank D.~Gaggero, E.~Del Nobile, and M.~Pierre for useful discussions. We thank K.~Kannike for constructive remarks on the first version of this paper. This project has received support by the exchange of researchers project ``The flavour of the invisible universe" funded by the Italian Ministry of Foreign Affairs and International Cooperation (MAECI) and by the Slovenian Slovenian Research Agency (bilateral grant n. BI-IT-18-20-002, SI18MO07). The work of F.D. is supported by the research grants: ``The Dark Universe: A Synergic Multi-messenger Approach'' number 2017X7X85K under the program PRIN 2017 funded by the Ministero dell'Istruzione, Universit\`a e della Ricerca (MIUR); ``New Theoretical Tools for Axion Cosmology'' under the Supporting TAlent in ReSearch@University of Padova (STARS@UNIPD); ``New Theoretical Tools to Look at the Invisible Universe'' funded by the University of Padua. F.D. is also supported by Istituto Nazionale di Fisica Nucleare (INFN) through the Theoretical Astroparticle Physics (TAsP) project. F.D. acknowledges support from the European Union's Horizon 2020 research and innovation programme under the Marie Sk\l odowska-Curie grant agreement No 860881-HIDDeN. N.~K. and A.~S. acknowledge the financial support from the Slovenian Research Agency (research core funding No. P1-0035 and J1-8137). This article is based upon work from COST Action CA16201 PARTICLEFACE supported by COST (European Cooperation in Science and Technology). A.~S. is supported by the Young Researchers Programme of the Slovenian Research Agency under the grant No. 50510, core funding grant P1-0035. This project has received support by the European Research Council (ERC) under the European Union's Horizon 2020 research and innovation programme under grant agreement 833280 (FLAY), and by the Swiss National Science Foundation (SNF) under contract 200021-175940.

\appendix

\begin{widetext}

%%%%%%%%
%%%%%%%%
%%%%%%%%
\section{Dark matter annihilation cross sections}
\label{app:AnnihilationXSec}
%%%%%%%%
%%%%%%%%
%%%%%%%%

In this Appendix, we derive the most general expressions for the DM annihilation cross section for each allowed channel. For a given LQ state defined by the LQ hypercharge $Y$ and weak isospin $T$, we express our results in terms of the parameters defined in Sec.~\ref{sec:lq-classification}
%%%%%%%%%%%%%%%%
\begin{equation}
  \lbrace m_S,\, m_\Delta,\, \lambda_i,\, y^L_{ij},\, y^R_{ij} \rbrace\,.
\end{equation}
%%%%%%%%%%%%%%%%

\paragraph*{(i)~\underline{$SS \to hh$}~:} Feynman diagrams for annihilation into Higgs boson pairs are shown in Fig.~\ref{fig:diagram-SShh}. The couplings in \eqref{eq:VBSM2} lead to the differential cross section
%%%%%%%%%%%%%%%
\begin{equation}
  \label{eq:SShh}
  \begin{split}   
 \left[\dfrac{\mathrm{d}\sigma}{\mathrm{d}\cos\theta}\right]_{SS \to h h}  & = \frac{1}{2} \dfrac{|\lambda_4|^2}{32 \pi s} \sqrt{\dfrac{1-4 m_h^2/s}{1-4 m_S^2/s}} \left| 1+\dfrac{3\, m_h^2}{s-m_h^2+ i m_h \Gamma_h} + \dfrac{\lambda_4\, v^2}{t-m_S^2}+\dfrac{\lambda_4\, v^2}{u-m_S^2}\right|^2\,,
  \end{split}
\end{equation}
%%%%%%%%%%%%%%%
where the Mandelstam variables are bound to satisfy $s+t+u=2(m_S^2+m_h^2)$, $\Gamma_h$ is the total Higgs decay width and the angle $\theta$ is related to $t$ via the relation
%%%%%%%%%%%%%%%%
\begin{equation}
  \begin{split}   
    t &= m_S^2 + m_h^2 -\frac{s}{2}\left(1-\sqrt{1-\frac{4m_S^2}{s}}\sqrt{1-\frac{4m_h^2}{s}}\cos\theta\right).
%    u(\cos\theta) &= m_S^2 + m_h^2 -\frac{s}{2}\left(1+\sqrt{1-\frac{4m_S^2}{s}}\sqrt{1-\frac{4m_h^2}{s}}\cos\theta\right).\\
  \end{split}
\end{equation}
%%%%%%%%%%%%%%%
The overall combinatorial factor $1/2$ accounts for two identical final state particles. If we take the non-relativistic limit $s \to 4m_S^2$, which is the leading contribution to DM freeze-out, and we stay away from the Higgs resonance at $m_S \simeq m_h /2$, we find 
\begin{equation}
  (\sigma v)_{SS \to hh} = \frac{|\lambda_4|^2}{16\pi m_S^2} \sqrt{1-\frac{m_h^2}{m_S^2}} \left|1+\frac{6 m_S^2}{m_h^2-4m_S^2} - \frac{\lambda_4 v^2}{m_h^2-2m_S^2}\right|^2.
\end{equation}

\begin{figure}[h]
\centering
	\includegraphics[width=.95\linewidth]{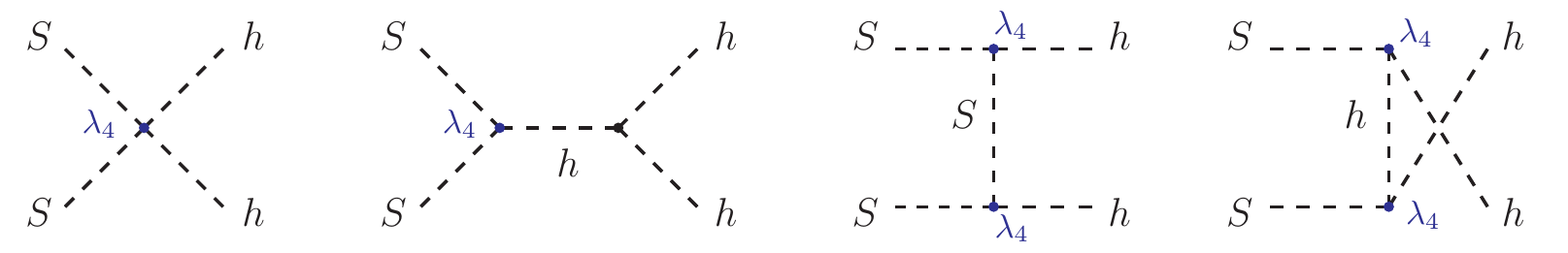}
\caption{\small \sl Tree-level contributions to the process $SS\to hh$.}
\label{fig:diagram-SShh}
\end{figure}

\paragraph*{(ii)~\underline{$SS \to \Delta \Delta^\ast$}~:} Annihilations to scalar leptoquarks are tree-level processes with diagrams shown in Fig.~\ref{fig:diagram-SSDD}. The total cross section for this process reads
\begin{equation}
  \label{eq:SSDD}
\left[\sigma\right]_{SS \to \Delta \Delta^\ast} = \dfrac{N_c (2T+1)}{16\pi s} \sqrt{ \dfrac{1-4 m_{\Delta}^2/s}{1-4 m_S^2/s} } \left| \lambda_3 + \dfrac{\lambda_4 \, \lambda_5 \, v^2}{s-m_h^2+i m_h \Gamma_h}\right|^2\,.
\end{equation}
where $N_c=3$ denotes the number of colors, and $2T+1$ accounts for the LQ weak-isospin multiplicity, i.e.~$T = 0$ for a weak singlet, $T =1/2$ for a doublet and $T=1$ for a triplet LQ. Away from the Higgs pole and in the non-relativistic limit we have the expression
\begin{equation}
  (\sigma v)_{SS \to \Delta \Delta^\ast} = \dfrac{N_c (2T+1)}{32 \pi m_S^2} \sqrt{1-\frac{m_\Delta^2}{m_S^2}} \left| \lambda_3 + \dfrac{\lambda_4 \, \lambda_5 \, v^2}{4 m_S^2-m_h^2} \right|^2\,.
\end{equation}
Note that this is the only annihilation process that depends on $\lambda_5$ coupling at tree-level.

\begin{figure}[h]
\centering
	\includegraphics[width=.49\linewidth]{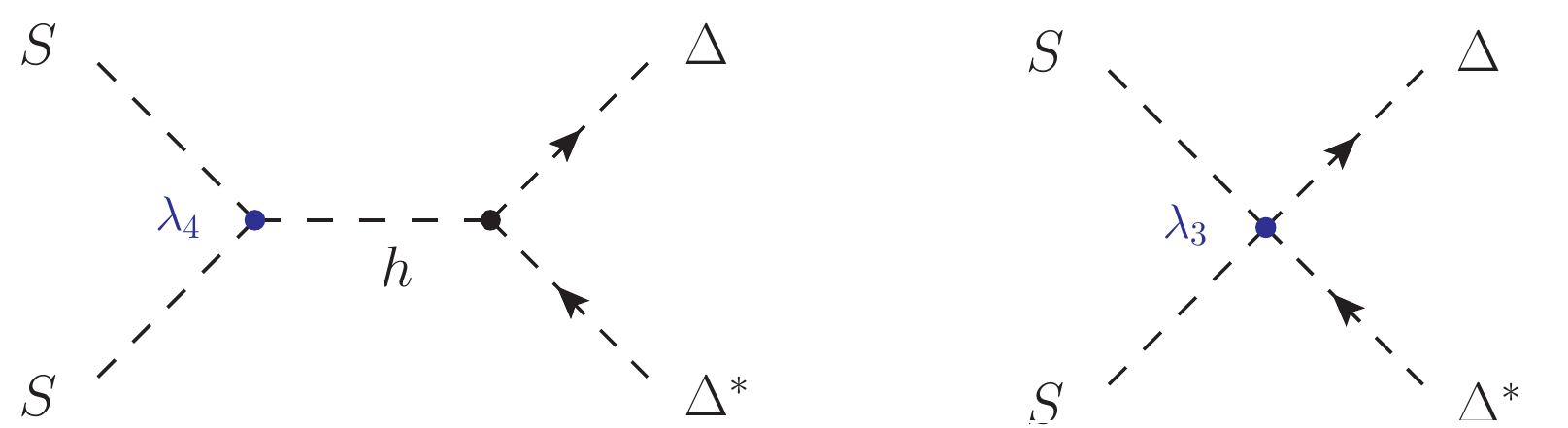}
\caption{\small \sl Tree-level contributions to the process $SS\to \Delta\Delta^\ast$.}
\label{fig:diagram-SSDD}
\end{figure}

\paragraph*{(iii)~\underline{$SS \to V_1V_2$}~:} DM can also annihilate to SM gauge bosons, and these processes can proceed either via an $s$-channel Higgs mediated contribution followed by a tree (loop) mediated $hV_1 V_2$ vertex for massive (massless) vector bosons, or via direct loop diagrams. The associated Feynman diagrams are depicted in Fig.~\ref{fig:diagram-SSVV}. We parameterize the $S S \to V_1 V_2$ loop amplitude as
%%%%%%%%%%%%%%%%
\begin{equation}
  \label{eq:DVVdef}
  \begin{split}
  \mathcal{A}_{V_1 V_2}^{\mathrm{loop}}&= -4 i\, (1+\delta_{V_{1}V_{2}})  \,D_{V_1 V_2}\left[ p_1 \cdot p_2\, g^{\mu\nu} - p_1^\nu p_2^\mu\right]\, \epsilon_{V_1}^{\mu*}(p_1,A) \epsilon_{V_2}^{\nu*}(p_2,B)\,T^{AB}\,,
  \end{split}
\end{equation}
where $D_{V_1 V_2}(s)$ are form-factors, and $\epsilon_{V}^\mu (p, A)$ denotes the $V$-boson polarization with momentum $p$ and index $A$. For gluons, $A=1,\ldots,8$ and
$T^{AB}= \delta^{AB}/2$, while for the electroweak bosons one should replace $T^{AB}$ by 1. The factor $\delta_{V_1 V_2}$ accounts for
identical particles in the final state, being $\delta_{VV}=1$ for $V\in \lbrace \gamma, g, Z\rbrace$ and $\delta_{V_1V_2}=0$
otherwise. The above gauge-invariant form is valid in the $m_{V_{1,2}} \ll m_\Delta$ limit~\footnote{In the effective theory limit, $\sqrt{s}, m_{V_{1,2}} \ll m_\Delta$, in such a way that the above amplitude
  corresponds to the effective Lagrangian $S^2 \sum_{(XY)} D_{(XY)} \,F^{(XY)}_{\mu\nu}  F^{(XY)\mu\nu}$ where $(XY) = (\gamma\gamma),(gg),(\gamma Z),(ZZ),(WW)$.}, a good approximation supported by the lower bounds on $m_\Delta$ determined by direct searches for LQs at the LHC~\cite{Angelescu:2018tyl}. The general expressions we have obtained
for $D_{V_1 V_2}$ are also reported in App.~\ref{app:formfactors}. By using the expressions defined above we can express the cross section for $SS\to V_1V_2$ in terms of $D_{V_1V_2}$ and of the Higgs coupling to vector bosons. We start by considering the $SS$ scattering into $WW$ and $ZZ$. In this case, the Higgs-mediated diagram in Fig.~\ref{fig:diagram-SSVV} appears already at tree-level
%%%%%%%%%%%%%%%
\begin{align}
\label{eq:sigma-WZ}
\sigma_{SS \to VV} = \dfrac{1}{4 \pi s} \sqrt{\dfrac{1-4 m_V^2/s}{1-4 m_S^2/s}} &\Bigg{\lbrace} \dfrac{1}{1+\delta_{V V}}\dfrac{\lambda_4^2 m_V^4}{(s-m_h^2)^2+m_h^2 \Gamma_h^2}\Bigg{[}2+\dfrac{(s-2m_V^2)^2}{4m_V^4}\Bigg{]}\nonumber\\
&+2 (1+\delta_{V V}) |D_{V V}^\mrm{SD}|^2 m_V^4 \Bigg{[}2+\dfrac{(s-2m_V^2)^2}{m_V^4}\Bigg{]}\\
&+12 \lambda_4\,  \mathrm{Re} \Bigg{[}\dfrac{D_{V V}^\mrm{SD} \, m_V^4}{s-m_h^2+i m_h \Gamma_h} \Bigg{]}\left(1-\dfrac{s}{2m_V^2}\right)\Bigg{\rbrace}\,,\nonumber
\end{align}
%%%%%%%%%%%%%%%
where $V=W,Z$, and $\delta_{VV}$ is such that $\delta_{ZZ}=1$ and $\delta_{W^+W^-}=0$. The leptoquark-loop contributions $D_{V V}^\mrm{SD}$ can be found in App.~\ref{app:formfactors}. For the remaining annihilation channels, namely $SS\to \gamma \gamma$, $SS \to gg$ and $SS\to \gamma Z$, the Higgs-mediated contribution appears only at one-loop level and is included in the definition of the $D_{V_1 V_2}$ form-factors. We obtained
%%%%%%%%%%%%%%%
\begin{align}
\sigma_{SS\to\gamma\gamma} &= \frac{1}{\pi} \frac{s \, |D_{\gamma\gamma}|^2}{\sqrt{1-4m_S^2/s}} \,,\\
\sigma_{SS\to g g} &= \frac{2}{\pi} \frac{s \, |D_{gg}|^2}{\sqrt{1-4m_S^2/s}} \,,\\
\sigma_{SS\to\gamma Z} &= \frac{1}{2\pi}\left(1-\dfrac{m_Z^2}{s}\right)^3 \frac{s \, |D_{\gamma Z}|^2}{\sqrt{1-4m_S^2/s}} \,,
\end{align}
with explicit expressions for $D_{\gamma\gamma}$, $D_{gg}$, and $D_{\gamma Z}$ reported in App.~\ref{app:formfactors}.

\begin{figure}[h]
\centering
	\includegraphics[width=.93\linewidth]{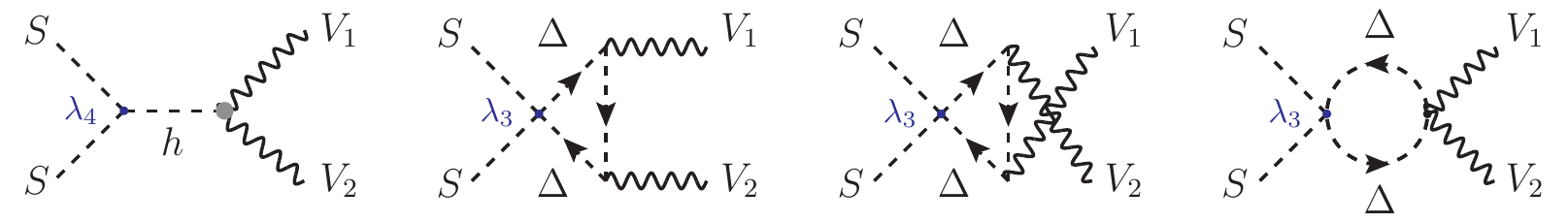}
\caption{\small \sl Tree- and one-loop diagrams for $SS\to V_1 V_2$ process, with $V_{1,2}$ a SM gauge boson. The gray blob in the first diagram appears at tree-level for $SS\to W W$ and $SS\to ZZ$, while the same couplings arises at one-loop level $SS\to \gamma\gamma$ and $SS\to \gamma Z$.}
\label{fig:diagram-SSVV}
\end{figure}

\paragraph*{(iv)~\underline{$SS \to f \bar{f}$}~:}  Finally, we derive the expression for the DM annihilation into fermions. Previous studies have only considered the tree-level Higgs-mediated contributions for these processes~\cite{Choi:2018stw}. Loop-induced ones can also be relevant for $f=\ell$ due to a chiral enhancement ($\propto m_q/m_\ell$) which overcomes the suppression by the small lepton Yukawas of the tree-level diagrams. The relevant diagrams for this process are depicted in Fig.~\ref{fig:dm-fermions}, which can proceed either via $\lambda_4\, S^2 |H|^2$ interaction followed by the $h\ell^+ \ell^-$ vertex (left panel), or via $\lambda_3 \, S^2 |\Delta|^2$ (right panel). 

\begin{figure}[h]
\centering
	\includegraphics[width=.57\linewidth]{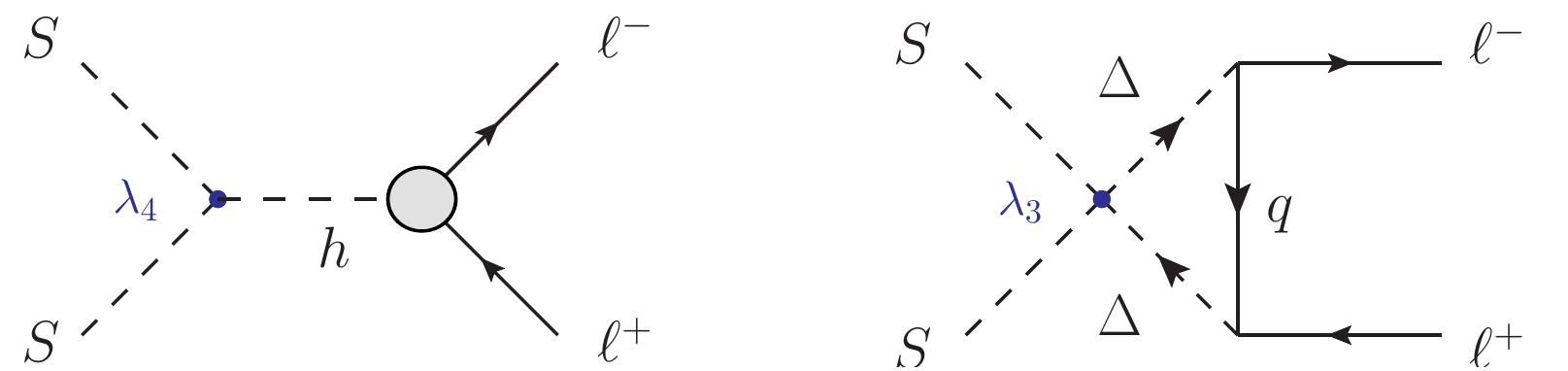}
\caption{\small \sl Leading contributions to the process $SS\to \ell^+\ell^-$. In the first diagram, the gray blob also includes the LQ loop modification of the $h \ell^+ \ell^-$ coupling. See text for details.}
\label{fig:dm-fermions}
\end{figure}

First, we discuss the contributions from the left diagram in Fig.~\ref{fig:dm-fermions}. These contributions amount to an effective modification of Higgs Yukawa coupling,
  \begin{equation}
    \label{eq:ModYuk}
    [y^\mathrm{eff}]_{\ell\ell} = \frac{\sqrt{2}  m_\ell}{v} + \frac{3v^2}{8\pi^2 m_\mrm{\Delta}^2} y_t (y_t^2-\lambda)\,y^R_{t \ell} y^{L*}_{t \ell}\,\log\frac{m_h}{m_\mrm{\Delta}}.    
  \end{equation}
where we have kept only the dominant contributions, which arise from top-quark loops, in the leading logarithm approximation. This contribution can be combined with the LQ loops depicted in the right diagram of Fig.~\ref{fig:dm-fermions}, which also induce chirality-enhanced contributions, in such a way that the total cross section reads
%%%%%%%%%%%%%%%
\begin{align}
\label{eq:SSll}
  \sigma_{SS \to \ell^+\ell^-}& = \dfrac{m_\ell^2}{8\pi} \dfrac{(1-4m_\ell^2/s)^{3/2}}{(1-4m_S^2/s)^{1/2}} \\
                                           &\phantom{=}\times\Bigg[ \Bigg|\dfrac{\lambda_4}{s-m_h^2+i m_h \Gamma_h} \dfrac{v\,\Re[y^{\mathrm{eff}}]_{\ell\ell}}{\sqrt{2}m_\ell}+\dfrac{\lambda_3}{16\pi^2 m_\Delta^2} \sum_{q} \dfrac{m_q}{m_\ell}\, \Re(y^L_{q\ell}\, y^{R*}_{q\ell}) G (y,x_q)\Bigg|^2 \nonumber\\
    &\phantom{=\times}+\left|\dfrac{\lambda_4}{s-m_h^2+i m_h \Gamma_h} \dfrac{v\,\Im[y^{\mathrm{eff}}]_{\ell\ell}}{\sqrt{2}m_\ell} +  \dfrac{\lambda_3}{16\pi^2 m_\Delta^2} \sum_{q} \dfrac{m_q}{m_\ell}\, \Im(y^L_{q\ell}\, y^{R*}_{q\ell}) G (y,x_q)\right|^2 \Bigg]\,,\nn
\end{align}
where we assume $m_\ell \ll \sqrt{s}$ and $m_\ell \ll m_q$, and we separate the real and imaginary parts of the Yukawa couplings which generate non-interfering scalar and pseudoscalar amplitudes, respectively. The loop-function $G(y,x_q)$, with $y=s/(4 m_\Delta^2)$ and $x_q\equiv m_q^2/m_\Delta^2$, is
%%%%%%%%%%%%%%%
\begin{equation}
G(y,x_q) \equiv C_0(0,0,4y;1,x_q,1)\,,
%G(s;m_f,m_{\Delta})\equiv C_0(0,0,s;m_\Delta^2,m_f^2,m_\Delta^2)\,,
\end{equation}
%%%%%%%%%%%%%%%

\noindent where $C_0$ stands for the three-point Passarino-Veltman function~\cite{tHooft:1978jhc}, with the same conventions as used in the {\sc Package-X} documentation~\cite{Patel:2015tea}.
Finally, note that in the case of $SS \to q \bar{q}$ the loop contributions are expected to be less important since the chirality factor would now be $m_{\ell}/m_q$. Therefore we approximate the cross section by its the tree-level contribution via the Higgs-portal
\begin{equation} 
\left[\sigma\right]_{SS \to q \bar{q}} =  \dfrac{N_c m_q^2}{8\pi} \frac{\left(1-4 m_q^2/s\right)^{3/2}}{\left(1-4 m_S^2/s\right)^{1/2}}  \left|\dfrac{\lambda_4}{s-m_h^2+i m_h \Gamma_h}\right|^2.
\end{equation}

%%%%%%%%
%%%%%%%%
%%%%%%%%
\section{One-loop results}
\label{app:loop}
%%%%%%%%
%%%%%%%%
%%%%%%%%

Within our framework, loop effects have a phenomenological relevance in several contexts. We collect in this Appendix one-loop results.

\subsection{LQ corrections to Lepton Yukawa couplings}

We begin with an estimate of the LQ contributions to the Higgs Yukawa coupling to leptons at one-loop. To this purpose, we integrate-out the LQs at tree-level and incorporate the leading-logarithm contribution from electroweak running~\cite{Feruglio:2018fxo} to the operator
\begin{equation}
Q_{\substack{eH \\ rs}} = |H|^2 \bar L_r H e_s  \ .
\end{equation}
The presence of this operator at the electroweak scale breaks the SM linear relation between Higgs Yukawas and lepton masses, thus inducing modification of the DM annihilation cross sections. Potentially large contributions with non-chiral couplings are only from $R_2$ and $S_1$ leptoquarks which generate at tree-level a non-chiral operator,
%%%%%%%%%%%%%%
\begin{equation}
Q_{\substack{lequ\\srpt}} = (\bar L_p^j e_r) \epsilon_{jk} (\bar Q_s^k u_t) \,.
\end{equation}
Closing the heavy-quark loop and attaching to it Higgs line(s) causes mixing into $Q_{\substack{eH \\ rs}}$ operator~\cite{1308.2627,1310.4838,1312.2014},
%%%%%%%%%%%%%
\begin{align}
    \label{eq:SM-RGE}
    C_{\substack{eH\\ rs}} (\mu_\mrm{EW}) &= \frac{12 y_t (y_t^2-\lambda)}{(4\pi)^2} \log \frac{\mu_\mrm{EW}}{\Lambda} \,C_{\substack{lequ\\rs33}} (\Lambda)\,,\\
    [Y_e]_{rs} (\mu_\mrm{EW}) &= [Y_e^\mrm{SM}]_{rs}(\mu_\mrm{EW}) -
                                \frac{6\lambda y_t v^2}{(4\pi)^2} \log \frac{\mu_\mrm{EW}}{\Lambda}\,C_{\substack{lequ\\sr33}} (\Lambda)\,.
\end{align}
%%%%%%%%%%%%  
\noindent where $\mu_\mrm{EW} \approx m_h$ denotes the electroweak scale, $\Lambda \approx m_\mrm{\Delta}$  is the matching scale where a LQ is integrated out, and we have kept only the leading-logarithm contributions in this expression. The dim-6 operator misaligns the Yukawa couplings relative to the lepton mass matrix $M_\ell$, such that the Yukawa coupling reads, in the mass basis, 
%%%%%%%%%%%%%%%%
\begin{equation}
     [y^\mathrm{eff}]_{\ell\ell} = \frac{\sqrt{2}  m_\ell}{v} + \frac{3v^2}{4\pi^2} y_t (y_t^2-\lambda) \log\frac{\mu_\mrm{EW}}{\Lambda} C_{\substack{lequ\\\ell\ell 33}} (\Lambda)\,.
\end{equation}
%%%%%%%%%%%%%%%

\noindent It is clear that only the dimension-6 operator contributes to the effective Yukawa coupling modification, whereas the dimension-4 Yukawa has been absorbed in the weak-to-mass basis rotation matrices.  We have neglected running of the lepton masses below scale $\mu_\mrm{EW}$. If the effective coefficient $C_{\substack{lequ}}$ has an imaginary part then we get also a pseudoscalar-type Yukawa coupling $\bar \ell \, i \gamma_5 \, \ell h$. There are two
  non-chiral LQ models that contribute to the $C_{lequ}(\Lambda)$
  coefficients: $R_2$ which is a $F=0$ LQ, and $S_1$ with
  $|F|=2$. Their tree-level matching relations are
  \begin{equation}
    \label{eq:Qlequ}
    C_{\substack{lequ\\ prst }}^{R_2}(\Lambda) = \frac{y^R_{u_s \ell_r} y^{L*}_{u_t \ell_p}}{2 m_\mrm{\Delta}^2},\qquad\quad C_{\substack{lequ\\ prst }}^{S_1}(\Lambda) = \frac{y^R_{u_t \ell_r} y^{L*}_{u_s \ell_p}}{2 m_\mrm{\Delta}^2}\,,
  \end{equation}
  where the matrices $y^{L(R)}$ are defined in Table~\ref{tab:lq-matching-l}. Since we consider only couplings to top quarks and leptons with equal flavour, the only nontrivial flavour combination is $prst = \ell \ell t t$ in both cases. Inserting the above expressions into $y^\mrm{eff}$ yields Eq.~\eqref{eq:ModYuk}.  The non-logarithmic contributions to this matching have been recently computed in Ref.~\cite{Crivellin:2020tsz,Crivellin:2020mjs}, see~also~\cite{Gherardi:2020det}.

\subsection{$SS\to V_1 V_2$ and $h\to V_1 V_2$ form factors}
\label{app:formfactors}

We provide the form factors $D_{V_1 V_2}(s)$ that appear in the DM annihilation cross sections reported in App.~\ref{app:AnnihilationXSec}. The short-distance~(SD) LQ loop contributions are represented by the last three diagrams in Fig.~\ref{fig:diagram-SSVV}:
\begin{align}
  \label{eq:DVVdirect}
  D_{\gamma\gamma}^\mrm{SD} &=\frac{\alpha_{\mathrm{em}} \lambda_3}{32 \pi m_\Delta^2}   \,N_c (2 T+1)\left[\frac{T (T+1)}{3}+Y^2\right]\mc{A}_0 (x_\Delta)   \,,  \\[0.4em]
  D_{gg}^\mrm{SD} &=  \frac{\alpha_3 \lambda_3}{32 \pi  m_\Delta^2} (2T+1)\,\mc{A}_0 (x_\Delta)\,,\\[0.4em]
  D_{\gamma Z}^\mrm{SD}    &=\frac{\alpha_\mathrm{em} \lambda_3}{16\pi s_W c_W m_\Delta^2} \,N_c (2 T+1)\left[c_W^2 \dfrac{T(T+1)}{3} -s_W^2 Y^2\right]\,\mc{A}_0(x_\Delta) \,,\\
  \label{eq:DVVdirectZZ}
  D_{ZZ}^\mrm{SD} &= \frac{\alpha_2 \lambda_3}{32 \pi  c_W^2 m_{\Delta}^2}\, N_c (2 T+1)   \left[c_W^4 \dfrac{T (T+1)}{3}+Y^2 s_W^4\right]\mc{A}_0(x_\Delta) \,,\\[0.4em]
  D_{WW}^\mrm{SD} &= \frac{\alpha_2 \lambda_3 L_2(\Delta) N_c}{16 \pi m_{\Delta}^2} \mc{A}_0(x_\Delta)\,,                               
\end{align}
%%%%%%%%%%%%%
where $x_\Delta \equiv {s}/{4 m_\Delta^2}$. In these expressions we
have set the final state masses $m_{V_{1,2}} = 0$ consistently with
the gauge-invariant expression~\eqref{eq:DVVdef}. The $SU(2)_L$ Dynkin index $L_2(\Delta)$ takes values $0,1/2,2$ in case when weak isospin of $\Delta$ is $T=0,1/2,1$, respectively. Another consistency
check that we find valid is equality
$D_{\gamma Z}^\mrm{SD} = 2 D_{\gamma\gamma}^\mrm{SD}$ in the limit
$s_W \to -1/\sqrt{2}$, $c_W\to 1/\sqrt{2}$, $g_2 \to 0$, and
$T_3 \to 0$, where both $Z$ and $A$ are massless and have identical
term in the covariant derivative~\eqref{eq:Dmu}.

For $SS\to ZZ$ and $SS\to WW$ there is an additional $s$-channel Higgs mediated diagram which interferes with the leptoquark loop contribution $D_{V_1 V_2}^\mrm{SD}$, as explicitly shown in Eq.~\eqref{eq:sigma-WZ}. For the form factors involving a massless boson we can absorb such $s$-channel Higgs contributions into $D_{V_1 V_2}$:
%%%%%%%%%%%%%
\begin{align}
  \label{eq:DVV}
  D_{\gamma\gamma} &=   D_{\gamma\gamma}^\mrm{SD}-\dfrac{\lambda_4\, v \, g_{h\gamma\gamma}}{8(s-m_h^2+i m_h \Gamma_h)}\,,\\[0.4em]
	\label{eq:Dgg}
  D_{gg} &=   D_{gg}^\mrm{SD} -\dfrac{\lambda_4\, v \, g_{hgg}}{8(s-m_h^2+i m_h \Gamma_h)}\,,\\[0.4em]
  D_{\gamma Z} &=  D_{\gamma Z}^\mrm{SD} -\dfrac{\lambda_4\, v \, g_{h\gamma Z}}{4(s-m_h^2+i m_h \Gamma_h)}\,.
\end{align}
%%%%%%%%%%%%%
The effective loop-induced couplings $g_{hV_1 V_2}$ of gauge bosons with the Higgs are defined in analogy with the form factors~\eqref{eq:DVVdef} of $SS \to V_1 V_2$:
\begin{equation}
  \label{eq:ghVVDef}
  \mc{A}\left[h(s) \to V_1(p_1,A) V_2 (p_2, B)\right] \equiv i g_{hV_1 V_2}(s) (1+\delta_{V_1 V_2})(p_1 \cdot p_2 g^{\mu\nu} - p_1^\nu p_2^\mu) \epsilon_{1\mu}^* \epsilon_{2\nu}^* T^{AB}\,.
\end{equation}
The above amplitude corresponds to effective Lagrangian $\mc{L}_{hV_1 V_2} = -(1/2)g_{hV_1 V_2} h V_{1\mu\nu} V_2^{\mu\nu}$. The Higgs to diboson decay widths are then
\begin{equation}
  \label{eq:widths}
  \begin{split}
    \Gamma(h \to \gamma\gamma) &= \frac{m_h^3 |g_{h\gamma \gamma}|^2 }{16\pi},\\
    \Gamma(h \to gg) &= \frac{m_h^3 |g_{hgg}|^2 }{8\pi},\\
    \Gamma(h \to \gamma Z) &= \frac{m_h^3 (1-m_Z^2/m_h^2)^3 |g_{h\gamma Z}|^2 }{32\pi}.
  \end{split}
\end{equation}
The width of $h \to \gamma \gamma$ has a factor of 2 relative to $h \to \gamma Z$ due
to identical particles, whereas $h \to gg$ has additional factor of 2
that stems from the gluon octet sum,
$\sum_{A,B} (\delta^{AB}/2)^2 = (N_c^2-1)/4$.  The couplings $g_{h V_1 V_2}$ in the SM
have been computed in~\cite{Djouadi:2005gi,1206.1082,1301.4694} and
consist of loop contributions dominantly from gauge bosons and top
quark. The leptoquark contributions are analogous to the
diagrams of $SS \to V_1 V_2$ process shown in the rightmost three
diagrams in Fig.~\ref{fig:diagram-SSVV}, where we have to replace
$SS \Delta \Delta^*$ with the $h\Delta \Delta^*$ vertex. This
similarity between the two processes allows us to identify
$g_{hV_1 V_2} = -\frac{2\lambda_5 v}{\lambda_3} D^\mrm{SD}_{V_1 V_2}$. The expressions for $g_{hV_1 V_2}$ are
%%%%%%%%%%%%%%
\begin{align}
  \label{eq:higgs-photon}
  g_{h\gamma\gamma} &= -\dfrac{\alpha_\mrm{em}}{4\pi v } \Bigg{[} \mathcal{A}_1(x_W)+ N_c Q_t^2 \mathcal{A}_{1/2}(x_t) \\
  &\qquad\qquad+ \frac{\lambda_5 v^2}{4 m_{\Delta}^2}  N_c (2T+1)[Y^2+T(T+1)/3]\, \mathcal{A}_0(x_{\Delta})\Bigg{]}  \,,\nonumber\\
  \label{eq:higgs-gg}
  g_{hgg} &=-\frac{\alpha_3}{4\pi v} \left[\mc{A}_{1/2} (x_t) + \frac{\lambda_5  v^2}{4m_\Delta^2}  (2T+1) \mc{A}_0(x_\Delta)\right]\,,\\             
  g_{h\gamma Z} &= \dfrac{\alpha_\mrm{em}}{2 \pi v s_W c_W} \Bigg{[} c_W^2 \mathcal{B}_1(1/x_W,\lambda_W)+ 2 N_c Q_t \left(T_{3t}-2 Q_ts^2_W \right)   \mathcal{B}_{1/2}(1/x_t,\lambda_t) \\
  &\qquad\qquad - \frac{\lambda_5 v^2}{4 m_\Delta^2} \,N_c (2 T+1)\left(c_W^2 \dfrac{T(T+1)}{3} -s_W^2 Y^2\right) \mc{A}_0(x_\Delta) \Bigg{]}.\nonumber
\end{align}
%%%%%%%%%%%%%%

\noindent where $x_i = s/(4 m_i^2)$ and $\lambda_i = (4 m_i^2)/m_Z^2$,
for $i \in \lbrace W,t,\Delta\rbrace$, and the loop-functions are
reported in App.~\ref{app:aux-functions}. In these expressions, we
have neglected the sub-leading contributions from light fermion
loops. In these expressions we can further simplify sums over the weak
isospin states of $\Delta$:
%%%%%%%%%%%%%%%%
\begin{align}
\sum_{T_3} Q_{T_3}^2 &= (2T+1)[Y^2+T(T+1)/3]\,,
\end{align}
%%%%%%%%%%%%%%%%
and, similarly,
%%%%%%%%%%%%%%%%
\begin{align}
  \sum_{T_3} Q_{T_3}(T_3-Q_{T_3}s^2_W)=\, (2T+1)\left[ -Y^2 s^2_W + \dfrac{T(T+1)}{3} c^2_W\right]\,,\\
  \sum_{T_3} (T_3-Q_{T_3}s^2_W)^2=\, (2 T+1)   \left[c_W^4 \dfrac{T (T+1)}{3}+Y^2 s_W^4\right]\,.
\end{align}
%%%%%%%%%%%%%%%%
  
\subsection{Auxiliary functions}
\label{app:aux-functions}

The loop functions of triangle diagrams with a massive scalar and two massless vectors attached are parameterized by
%%%%%%%%%%%%%%%%
\begin{align}
  \label{eq:A0}
    \mathcal{A}_0(x) &= - (x-f(x))x^{-2}\,,\\[0.45em]
    \mathcal{A}_{1/2}(x) &= 2 \left[x+(x-1)\,f(x)\right] x^{-2}\,,\\[0.45em]
  \mathcal{A}_{1}(x) &= - \left[2x^2+3x+3(2x-1)\,f(x)\right] x^{-2}\,,\\
  & \nonumber
\end{align}
with
\begin{align}
f(x) &= \begin{cases}
\mathrm{arcsin}^2 \sqrt{x}\,,  &\hspace*{4.8em} x \leq 1 \\
-\dfrac{1}{4}\left( \log\dfrac{1+\sqrt{1-x^{-1}}}{1-\sqrt{1-x^{-1}}}-i \pi\right)^2\,, &\hspace*{4.8em} x >1
\end{cases},
\end{align}
where the indices $\lbrace 0,1/2,1\rbrace$ denote contribution of scalar, fermion, or vector running in the loop, respectively.
For the case when one of the external vectors is massive the above functions generalize to
\begin{align}
  \mc{B}_0(x,y) &= I_1(x,y)\,,\\[0.45em]
  \mc{B}_{1/2}(x,y) &= I_1(x,y)-I_2(x,y)\,,\\[0.45em]
    \mc{B}_{1}(x,y) & =4(3-\tan^2\theta_W)I_2(x,y)+[(1+2 x^{-1})\tan^2\theta_W-(5+2 x^{-1})]I_1(x,y)\,.
\end{align}
%%%%%%%%%%%%%%%
with the auxiliary functions defined by
%%%%%%%%%%%%%%%%
\begin{align}
I_1(x,y)&=\dfrac{xy}{2(x-y)}+\dfrac{x^2y^2}{2(x-y)^2}\left[f(x^{-1})-f(y^{-1})\right]+ \dfrac{x^2 y}{(x-y)^2}\left[g(x^{-1})-g(y^{-1})\right]\,,\\[0.45em]
I_2(x,y)&=-\dfrac{x y}{2(x-y)}\left[f(x^{-1})-f(y^{-1})\right]\,,
\end{align}
%%%%%%%%%%%%%%%%
and
\begin{equation}
g(x) = \begin{cases}
\sqrt{x^{-1}-1}\,\mathrm{arcsin} \sqrt{x}\,, &\qquad x \geq 1 \\ 
\dfrac{\sqrt{1-x^{-1}}}{2}\left( \log\dfrac{1+\sqrt{1-x^{-1}}}{1-\sqrt{1-x^{-1}}}-i \pi\right)^2\,,  &\qquad x <1
\end{cases}.
\end{equation}
%%%%%%%%%%%%%%%%
In the massless limit the $\mc{B}$ functions reduce to $\mc{A}$ functions: $\mc{B}_0 (1/x,\infty) = \mc{A}_0 (x)/2$,  $\mc{B}_{1/2} (1/x,\infty) = -\mc{A}_{1/2}(x)/4$.

\end{widetext}

%%%%%%%%
%%%% BIBLIOGRAPHY
%%%%%%%%

\end{document}